# Physics-Grounded Simulation Enables Zero-Shot Molecular and Reaction Analysis in Scanning Tunneling Microscopy

*Yuanqin He,[1] Zhiyu Wang,[2] Yang Bao,[3] Tao Lin,[2]* and Hai Bi[1]**

[1] Jihua Laboratory, Foshan, P. R. China

[2] College of New Materials and New Energies, Shenzhen Technology University, Shenzhen, P. R. China

[3] State Key Laboratory of Luminescence Science and Technology, Changchun Institute of Optics, Fine Mechanics and Physics, Chinese Academy of Sciences, Changchun, P. R. China

* e-mails: lintao@sztu.edu.cn; bihai@jihualab.ac.cn

## Abstract

Quantitative identification of molecular products and reaction pathways from scanning tunneling microscopy images is difficult when labeled images are unavailable for a new chemical system. We develop a simulation-assisted framework for zero-shot analysis that requires only precursor information for a new target system, with no experimental images or annotations from that system used during model training. The framework couples two functional components: **NanoSim** generates physics-grounded synthetic images from predicted reaction products and sampled surface scenes, while **NanoAnalysis** translates molecular detections into product populations, spatial arrangements, and mechanistic interpretations. Applied to a representative on-surface reaction, the detector achieves a micro-averaged F1 score of 0.92 on experimental images. Analysis of the detected molecules reveals annealing-dependent population changes and reaction-site preferences consistent with experimental observations. Beyond this capability, its modular organization allows heterogeneous data from other systems to be used to improve each component independently, providing a scalable route for continued generalization.

## Introduction

On-surface synthesis enables the bottom-up construction of diverse atomically precise nanoarchitectures, such as metal-organic networks, graphene nanoribbons, and covalent networks, providing promising routes toward nanoscale electronics, sensing, and quantum technologies.[1–9] Scanning tunneling microscopy (STM) is among the most widely used tools for studying these surface-confined architectures at the single-molecule level.[10] However, quantitative STM interpretation remains challenging because STM contrast reflects the coupled physics of tunneling, tip-sample interaction, and imaging conditions rather than molecular structure alone.[11–13] Reaction and surface heterogeneity further complicate the connection between STM images and chemical identity or pathway information.[14,15] Consequently, STM interpretation is still largely manual and difficult to scale to large datasets or system-level reaction analysis.

Recent advances in machine learning (ML) have enabled a range of automated surface-characterization tasks in STM, particularly image enhancement and molecular localization.[16–23] In parallel, ML methods have also been integrated with microscope control for automated tip conditioning, image acquisition, and autonomous exploration of surfaces.[24–29] More recent studies have extended these approaches toward molecular manipulation, reaction monitoring, and bond-selective transformations.[30–34] Despite these advances, most recognition pipelines still require experimental annotations from the target system and focus on individual species, making them difficult to extend to new molecular systems or higher-level reaction analysis.

Simulation offers a promising route to reducing the dependence on experimentally labeled data. Physically based approaches, including molecular dynamics, density functional theory, and electronic-structure-based imaging models, have been used to generate STM and related scanning probe images for known structures.[35–41] More recently, data-driven approaches have been explored to convert idealized simulations into experimental-like scanning probe microscopy (SPM)/STM images to reduce the simulation-experiment domain gap.[42–45] These studies demonstrate the potential of simulated data for STM analysis. However, existing approaches often remain centered on predefined structures or specific molecular systems, with limited consideration of the coupled chemical, surface, and imaging factors, restricting their extension to new molecules, surfaces, or acquisition settings.

In this work, we introduce a simulation-assisted framework for zero-shot STM analysis that requires only precursor information for a new target system, without target-system STM data or annotations, thus addressing the generalization limits of existing workflows. This capability is enabled by a simulation-assisted strategy that integrates two complementary components: **NanoSim**, which generates physics-grounded synthetic images from reaction, surface, and imaging priors, and **NanoAnalysis**, which derives perception-level predictions, system-level observables, and mechanistic insights from experimental images. In the representative BP system, the framework achieved a micro-F1 of up to 0.92 and enabled quantitative analysis of product populations, pathways, and annealing-dependent evolution. Its physics-grounded design further allows heterogeneous data across molecular systems and imaging conditions to improve individual modules independently, providing a scalable route for continued generalization.

## Results

### Overview of the Framework

The proposed framework comprises two components: **NanoSim**, which generates physics-grounded synthetic STM data from a precursor structure, and **NanoAnalysis**, which applies models trained on these data to experimental images (Figure 1). Because no experimental images or molecular annotations from the target system are used, the framework is zero-shot with respect to that system. To evaluate the framework, we selected 2-bromo-1,10-phenanthroline (BP) on Cu(111) as a model system, which undergoes debromination and dehydrogenation. Debromination is highly selective, whereas dehydrogenation follows multiple competing pathways, providing a controlled yet diverse testbed for selective chemistry and pathway heterogeneity. Additional evaluations on other molecular systems are provided in the Supplementary Information section S4.

**NanoSim** generates synthetic data across controlled levels of complexity and scale, which are subsequently used to train perception models for application to real STM images. It addresses the distribution gap between simulated and experimental images by explicitly encoding the governing physical principles of on-surface synthesis and STM imaging through three coordinated modules—a Physical layer, a Behavior layer, and a Rendering layer—each targeting a key stage of the nanoscale simulation workflow.

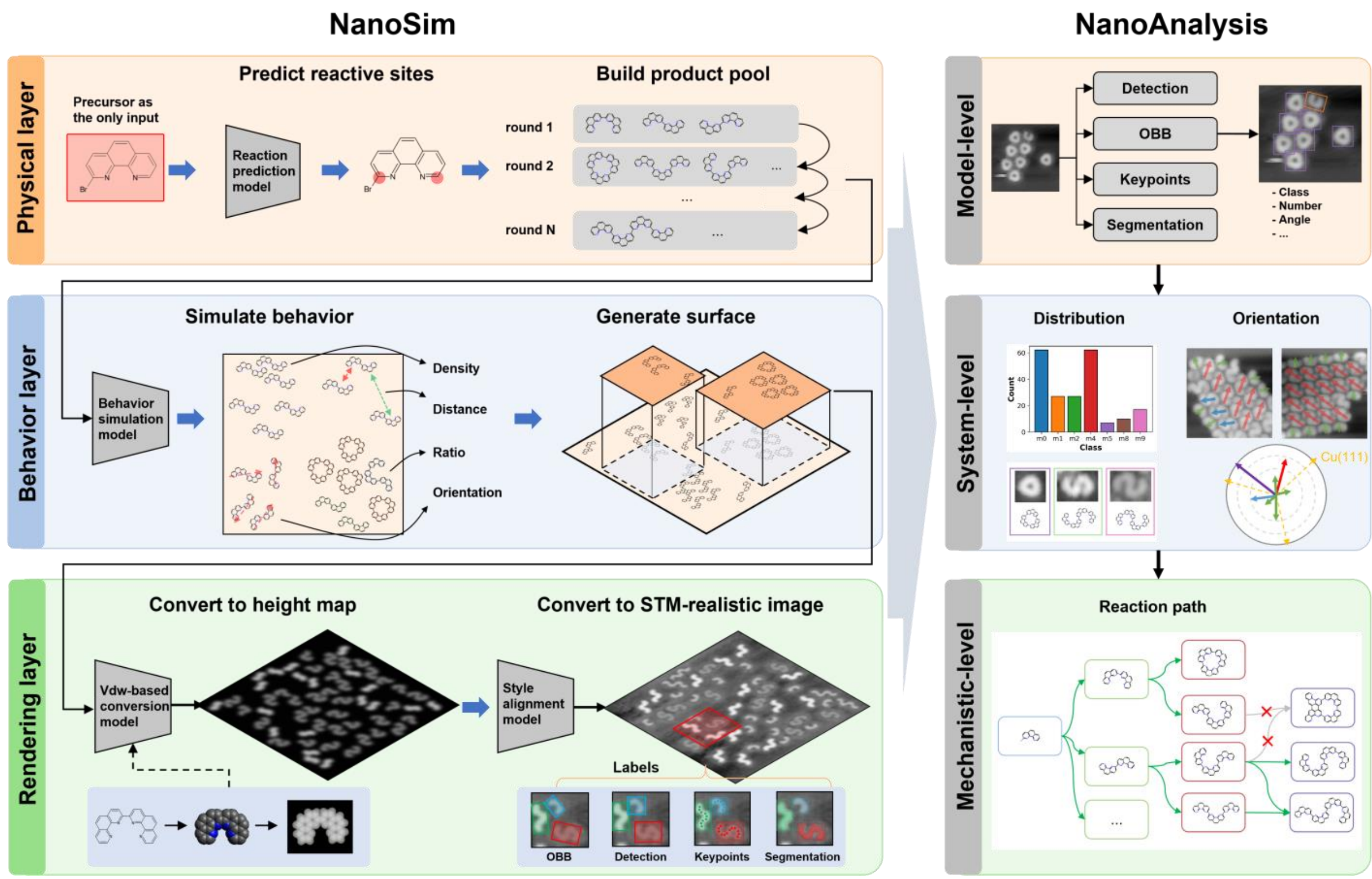

**Figure 1.** Overall scheme of the proposed framework, consisting of **NanoSim** and **NanoAnalysis**, for zero-shot analysis of on-surface reaction products from STM images.

The Physical layer constructs a reaction-guided candidate product space from the precursor structure. The reaction-prediction model[46] encodes chemical priors learned from reaction data, directing product generation toward confidence-ranked reactive sites rather than exhaustive enumeration. Top-ranked reactive sites are iteratively converted into candidate products, followed by surface-specific diversification and post-processing. For BP, the most probable site is used at each step and the reaction depth is set to three rounds unless otherwise noted. This procedure retains chemically relevant products while limiting the number of classes presented to downstream models (Supplementary Fig. 1).

The Behavior layer constructs physically plausible surface scenes from the predicted products by varying molecular number density, minimum edge-to-edge separation, relative abundance, and in-plane orientation. These interpretable parameters encode coarse physical priors on surface coverage, steric exclusion, scene composition, and orientational order. By enforcing these priors during molecular placement, the layer grounds scene generation in explicit physical constraints rather than unconstrained random sampling. This physics-grounded abstraction narrows the otherwise vast configurational space without requiring atomistically exhaustive surface modeling.

The Rendering layer converts each surface configuration into an STM-like image through two complementary stages. First, a vdW-sphere height map provides a scalable geometric proxy that preserves the projected molecular footprint and relative protrusions. This proxy captures structural geometry but does not replace bias-dependent electronic-structure STM simulation. Second, a Pix2Pix-based style-alignment module learns aggregate contrast and texture statistics from paired STM data for non-target molecular systems.[47,48] This hybrid strategy maintains physical grounding, narrows the simulation-experiment domain gap, and supports efficient large-scale synthetic-data generation. None of the molecular species evaluated here, including those in the Supplementary Information, were included in the Pix2Pix training set, thereby ensuring zero-shot evaluation.

**NanoAnalysis** organizes interpretation into three hierarchical levels. The perception level predicts molecular instances from synthetic supervision. The system level quantifies product abundance, density, orientation, intermolecular distance, and treatment-dependent population changes. The mechanistic level uses these observables to infer reaction-site selectivity, isomer preferences, and reaction pathways. Detailed procedures are provided in Supplementary Information Section S1.

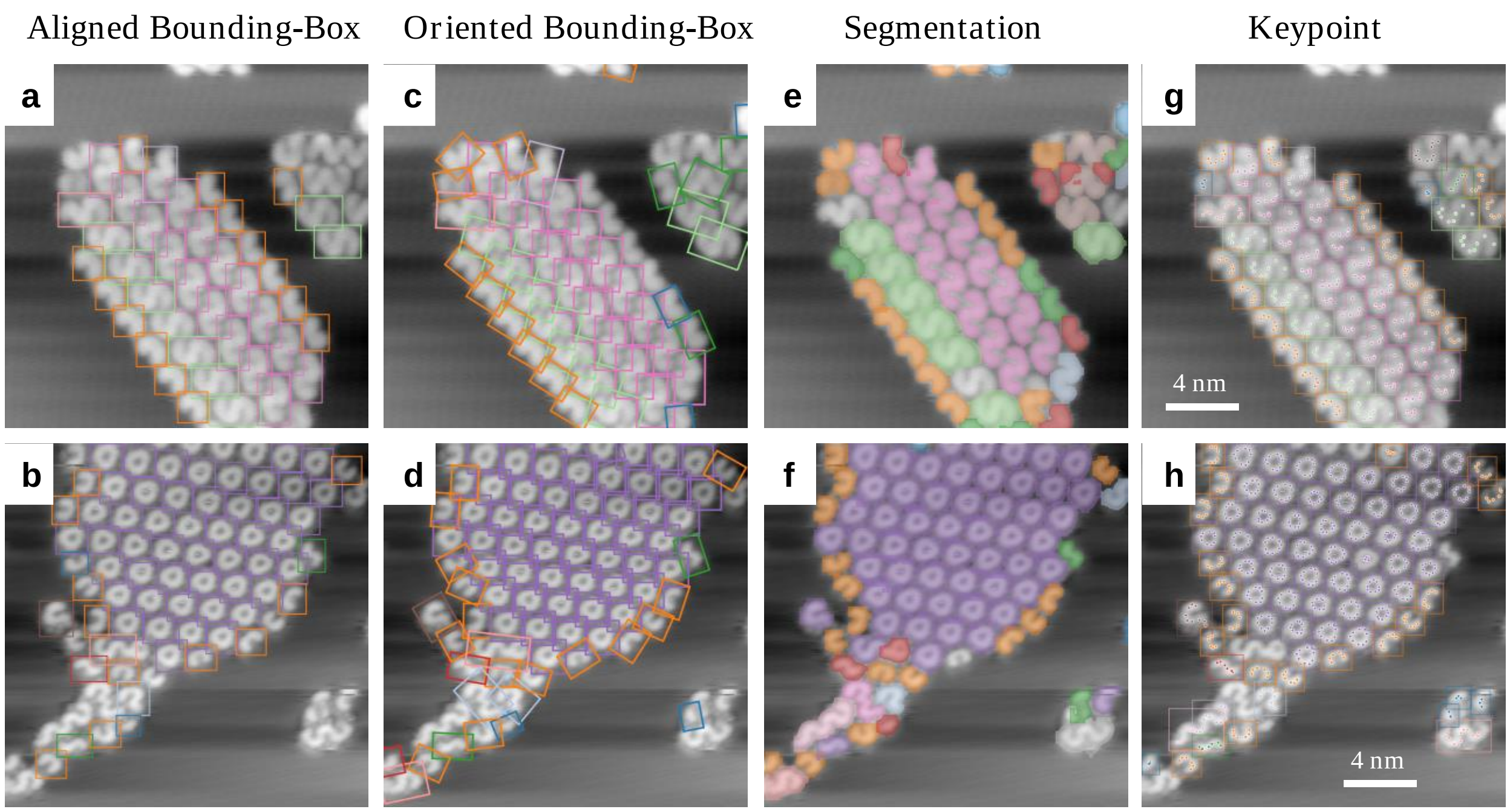

**Figure 2.** Representative results from the perception level of **NanoAnalysis** on experimental STM images: (a-b) axis-aligned bounding-box detection, (c-d) oriented bounding-box (OBB) detection, (e-f) instance segmentation, and (g-h) keypoint localization. Scanning parameters: U=-700 mV, I=100 pA.

At the perception level, base models are trained on the dense labels generated by **NanoSim**. Since **NanoSim** provides dense annotations at multiple semantic resolutions, it supports a broad range of perception tasks, including axis-aligned bounding-box detection, oriented bounding-box (OBB) detection, instance segmentation, and keypoint localization (see Supplementary Information Section S1.3 for training details).[49] Predictions on experimental images were visually consistent with the corresponding molecular structures across all four tasks, indicating reliable performance from coarse localization to fine-grained structural analysis (Figure 2). On a manually annotated test set of 10 real STM images (see Supplementary Information Section S1.4 for annotation and evaluation details), the OBB model trained on NanoSim data achieved a micro-F1 of approximately 0.90, with a precision of 0.92 and a recall of 0.86. These results demonstrate effective transfer from simulated to experimental images and provide reliable molecule-level inputs for higher-level **NanoAnalysis**.

At the system level, model outputs are converted into quantitative observables, including product abundance, orientation, and intermolecular-spacing statistics. We apply this analysis to 40 STM images acquired after annealing at 140 °C for 30 min. Figure 3a reports the abundances of products m0–m14; together with their confidence scores, this distribution helps distinguish likely products from rare or uncertain assignments requiring manual inspection. Representative structures agree with the observed STM contrasts (Figure 3b), supporting the assignments. Angular distributions reveal broad orientations for m1 but stronger substrate alignment for m4, m8, and m9, particularly m8 and m9 (Figure 3c). These distributions describe orientational preferences rather than complete adsorption registries. Nearest-neighbor analysis reveals pronounced self-aggregation of m4, m8 and m9, with m1 frequently distributed around the resulting m4-rich regions. Tetramers occur less often near m4, consistent with m4 being a terminal product that does not undergo further reaction.

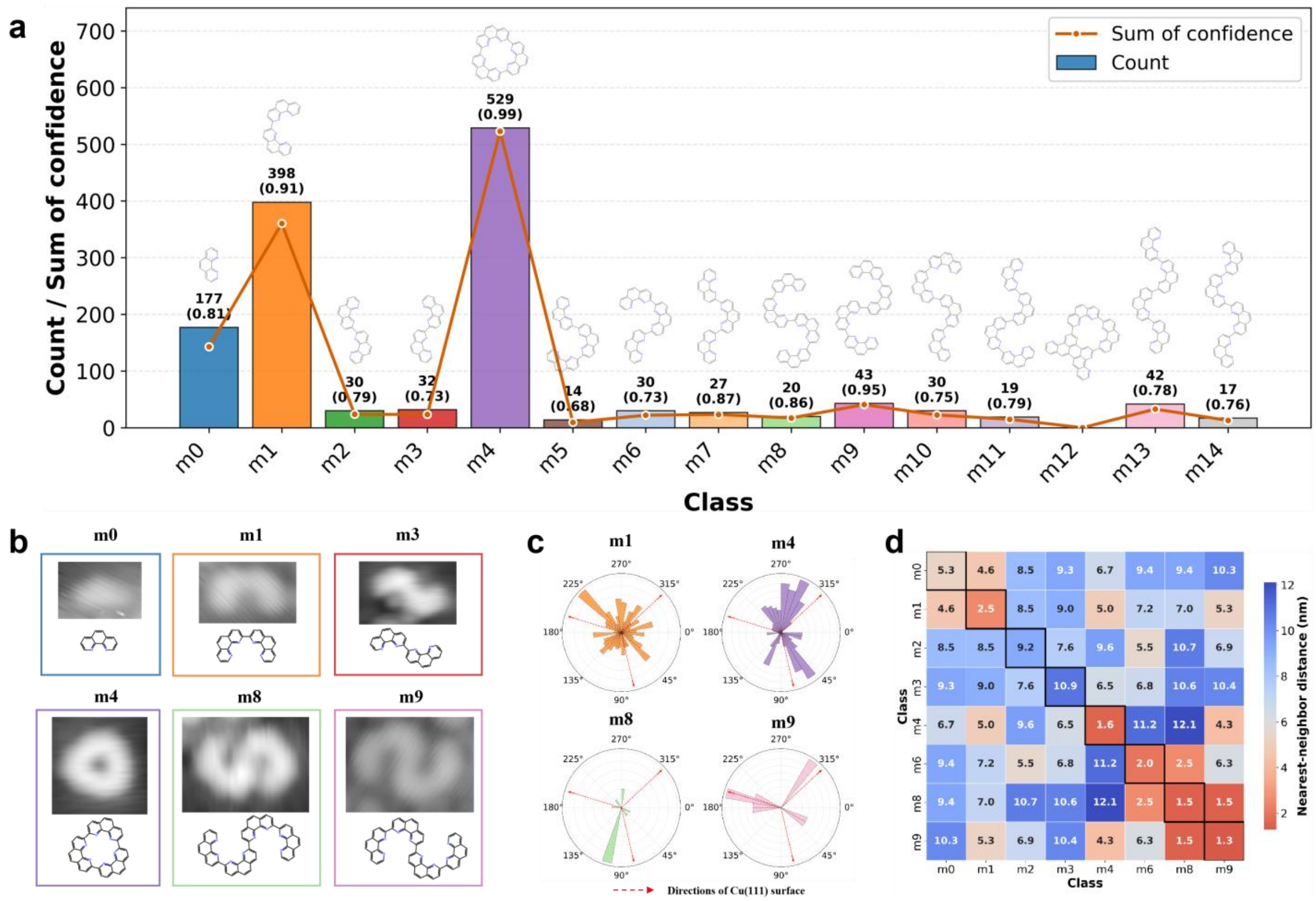

**Figure 3.** System-level analysis of **NanoAnalysis**. Molecule-level recognition outputs are converted into chemical observables: (a) product-population statistics, (b) representative product assignments with STM images, (c) angular orientation distributions, and (d) intermolecular spacing relationships.

At the mechanistic level, **NanoAnalysis** combines model predictions with experimental observations to interpret the multistep surface reaction. Figure 4a overlays measured product populations on the **NanoSim**-predicted reaction network in a Sankey diagram, with arrow width and color intensity representing population. For visualization purposes, product populations with multiple possible precursor channels were divided equally among the corresponding flows. This allocation reflects an assumption rather than resolved pathway-specific fluxes. Under this equal-allocation assumption, the m0-to-m1-to-m4 sequence appears as the most prominent flow. The predicted network contains no outgoing transformations from m4, and manual inspection identified no downstream products, supporting its assignment as a terminal product. In contrast, later-stage products m8, m9, m10, m11, m13, and m14 show broadly comparable populations, whereas m12 is not observed experimentally, likely because steric hindrance suppresses its formation. The majority of the dimer products adopt the cis configuration rather than the trans configuration, probably due to the template effect of Cu–N coordination.[50,51] The cis configuration further facilitates the formation of m4 in the subsequent reaction round.

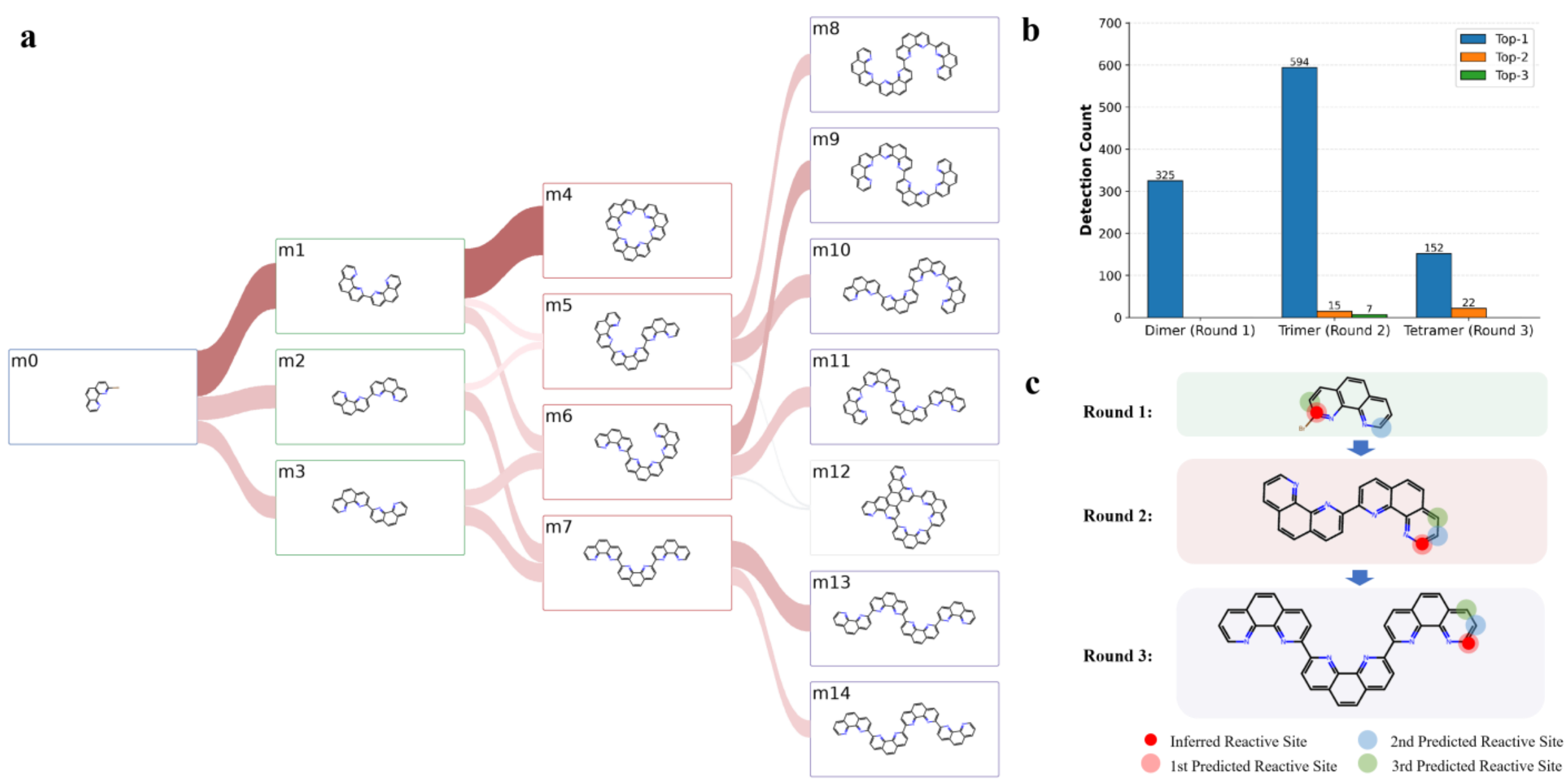

**Figure 4.** Mechanistic analysis using **NanoAnalysis** results: (a) Sankey representation of predicted transformations and product populations, (b) distribution of detected products within the Top-1 to Top-3 product space, and (c) predicted and experimentally inferred reaction sites.

**NanoAnalysis** can further examine reaction-site selectivity by mapping experimentally observed products back to their predicted precursor transformations. To quantify this selectivity, we constructed a Top-1 to Top-3 product space. Most experimentally detected products across the monomeric, dimeric, trimeric, and tetrameric stages matched the Top-1 predictions, with only a small fraction corresponding to Top-2 or Top-3 predictions (Figure 4b; corresponding product structures are provided in the Supplementary Information Section S2.1). The site-level correspondence is further illustrated in Figure 4c. For a representative molecule at each stage, the three highest-ranked predicted sites are shown in light red, green, and blue, while red dots mark the structural changes inferred from the assigned experimental products. Across the three reaction rounds, products consistent with the top-ranked predicted transformations were identified among the experimentally detected structures. Round 1 shows the expected selectivity for debromination, whereas in later rounds both prediction and experiment favor H atoms adjacent to N atoms, indicating that the model captures the key site-selectivity feature of the pathway. This preference is consistent with previous DFT calculations identifying the pyridine ortho position as the most reactive among the ortho-, meta-, and para-positions,[52] as well as with reported preferential ortho dehydrogenation on Cu(111).[53]

Beyond pathway reconstruction, we examined annealing-dependent population changes using 20, 40, and 40 STM images acquired before annealing and after 20 and 30 min at 140 °C, respectively (Figure 5; see Supplementary Information Section S2.4.2 for details).[54] Before annealing, m0 dominated the population (88%), with 11% m1 and negligible m4. After 20 min, m0 decreased to 12%, while m1 and m4 increased to 53% and 12%, respectively, indicating intermediate formation and the onset of m4 production. After 30 min, m4 increased further to 44%, accompanied by decreases in m0 (10%) and m1 (29%). Occasional confusion between m0 and m1 due to their similar STM appearances did not alter the overall population trend (Figure 5b). The remaining population comprised other lower-abundance products.

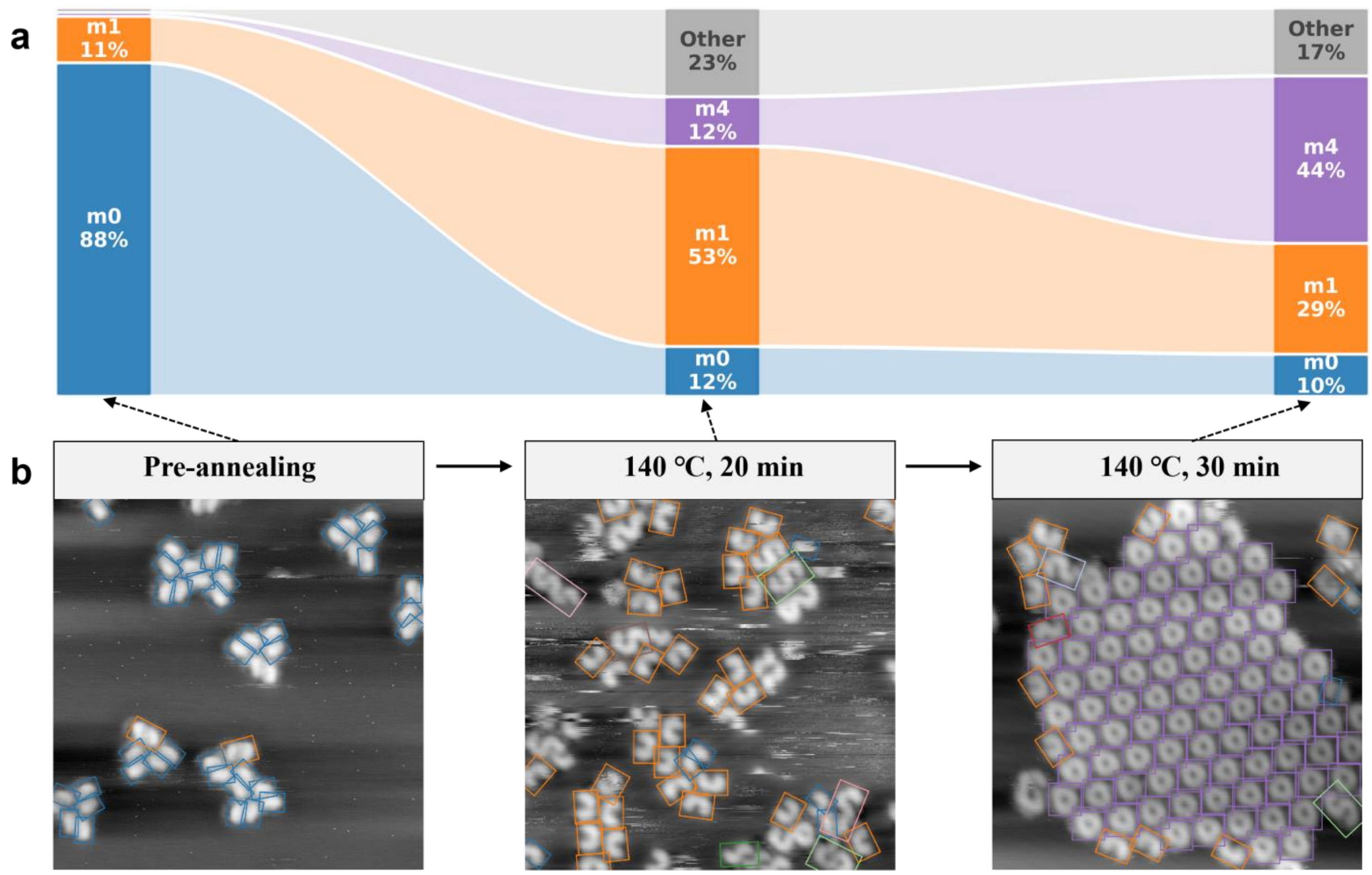

**Figure 5.** Evolution of BP product populations before annealing and after annealing at 140 °C for 20 and 30 min: (a) population flow and (b-d) representative STM images.

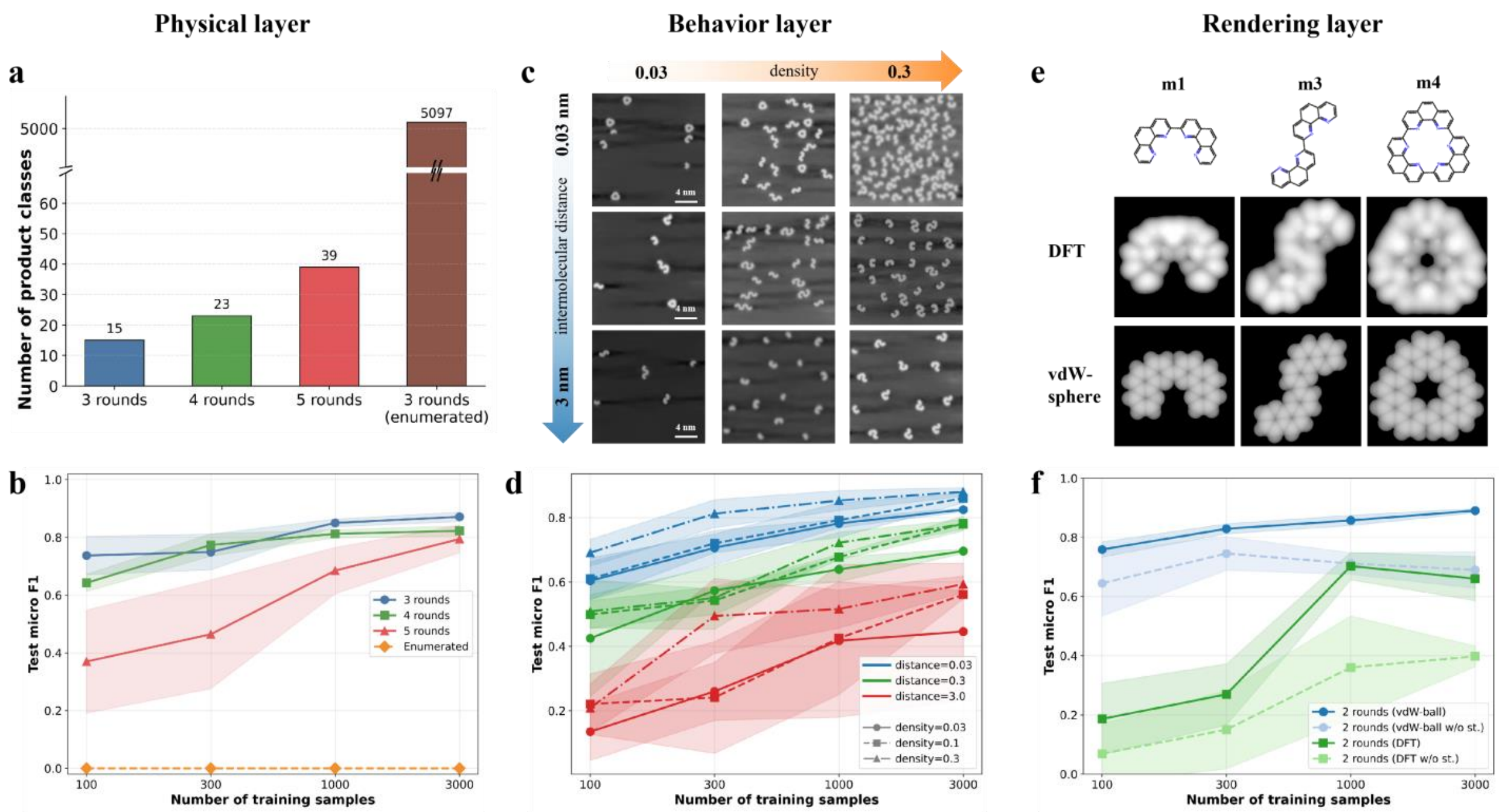

**Figure 6.** Ablation of **NanoSim** components: (a-b) reaction-guided product selection, (c-d) surface-scene sampling, and (e-f) vdW- versus DFT-based rendering.

## Key Components

To compare the contributions of individual **NanoSim** modules, we evaluated each design choice across multiple synthetic training-set sizes; detailed settings and ablation results are provided in the Supplementary Information Section S2.

First, we demonstrated that constraining the candidate product space is a key factor controlling downstream recognition performance (Figure 6a,b). Exhaustive enumeration generated over 5000 candidate product classes and yielded no meaningful recognition performance (micro-F1 = 0), showing that an excessively large product space obscures the relevant detection features. Reaction-prediction-based candidate spaces were much more compact: the three-, four-, and five-round settings generated 15, 23, and 39 candidate product classes, respectively. The three- and four-round product spaces gave comparable results, both exceeding a micro-F1 of 0.8 with 1000 synthetic training samples, whereas the five-round product space reduced performance by approximately 0.3. Because later-stage species are typically rare in the experimental test set, their inclusion enlarges the candidate space and increases recognition difficulty while contributing negligibly to the overall micro-F1. Detailed performance metrics are provided in Supplementary Information Section S2.1.

Second, we demonstrated that the behavior abstraction module strongly influences model performance (Figure 6c,d). Here, we use intermolecular distance and molecular density as representative parameters (see the Supplementary Information Section S1.2 for parameter definitions). Overall, the 3 nm spacing performed worst, with micro-F1 values below 0.6, whereas the 0.03 and 0.1 nm spacing settings both exceeded 0.8 with 3000 training samples. Recognition performance generally improved with increasing molecular density, as denser synthetic images provide more molecular instances and thus stronger training signals (Figure 6d).

Third, we compared the hybrid rendering module with DFT-based STM simulation (Figure 6e,f). Both approaches preserved similar molecular envelopes but differed in intramolecular contrast. With identical style alignment, vdW-sphere rendering reached a micro-F1 near 0.8, whereas DFT-based rendering plateaued near 0.6 after 1000 samples, likely because the former supports broader augmentation and greater synthetic-data diversity. Removing the style-alignment module substantially reduced performance (dashed lines), confirming its importance to the rendering strategy. It is worth noting that because three-round products were impractical for DFT simulation, this comparison used the two-round product space. Representative detection errors and framework limitations are discussed in Supplementary Information Section S3.

**Scaling Behavior**

Figure 7a evaluates the performance-time trade-off as the synthetic-data volume and YOLO11 model size increase. Increasing the training set from 100 to 100000 images and model capacity from 2.4 M to 55.7 M parameters raised the test micro-F1 from 0.52 to 0.92. Importantly, the framework achieved a micro-F1 close to 0.8 within 20 min and approximately 0.85 within 1 h. By contrast, DFT-based image generation for monomers, dimers and trimers was estimated to require over 200 h. Although enhanced cluster capacity could reduce this time, the computational cost remains substantially higher than that of generation via the framework. These results support efficient scaling of synthetic-data generation and downstream recognition.

Experimental datasets in surface chemistry are heterogeneous across molecular systems, instruments, and imaging conditions, limiting their scalable reuse. Our modular design allows each dataset to improve the relevant pipeline component independently, enabling effective use of diverse data sources.

Figure 7b evaluates scaling of the reaction-prediction module. Over three reaction rounds, the Top-1, Top-2, and Top-3 settings used the top one, two, and three predicted reaction sites and generated 15, 49, and 59 product classes, respectively. Overall, the Top-1 setting consistently outperformed the Top-2 and Top-3 settings and substantially outperformed exhaustive enumeration, demonstrating the benefit of a compact, chemically relevant product space. The Top-1 advantage narrowed from approximately 0.6 at 100 training samples to less than 0.1 at 3000 samples, indicating that additional data partially offset the complexity introduced by less certain reaction-site candidates. Because reaction-site prediction is not specific to a single STM system, this module can benefit directly from broader reaction datasets and improved predictive models.

Figure 7c examines the effect of Pix2Pix training-data volume on downstream recognition. Increasing the heterogeneous STM dataset from 50 to approximately 660 images raised the downstream micro-F1 from 0.70 to over 0.87 with 3000 synthetic training samples, consistent with the representative converted images (insets in Figure 7c). Since these STM data contain images from different molecular systems and experimental setups, these results show that cross-system STM images can be pooled within the style-alignment module to improve downstream recognition.

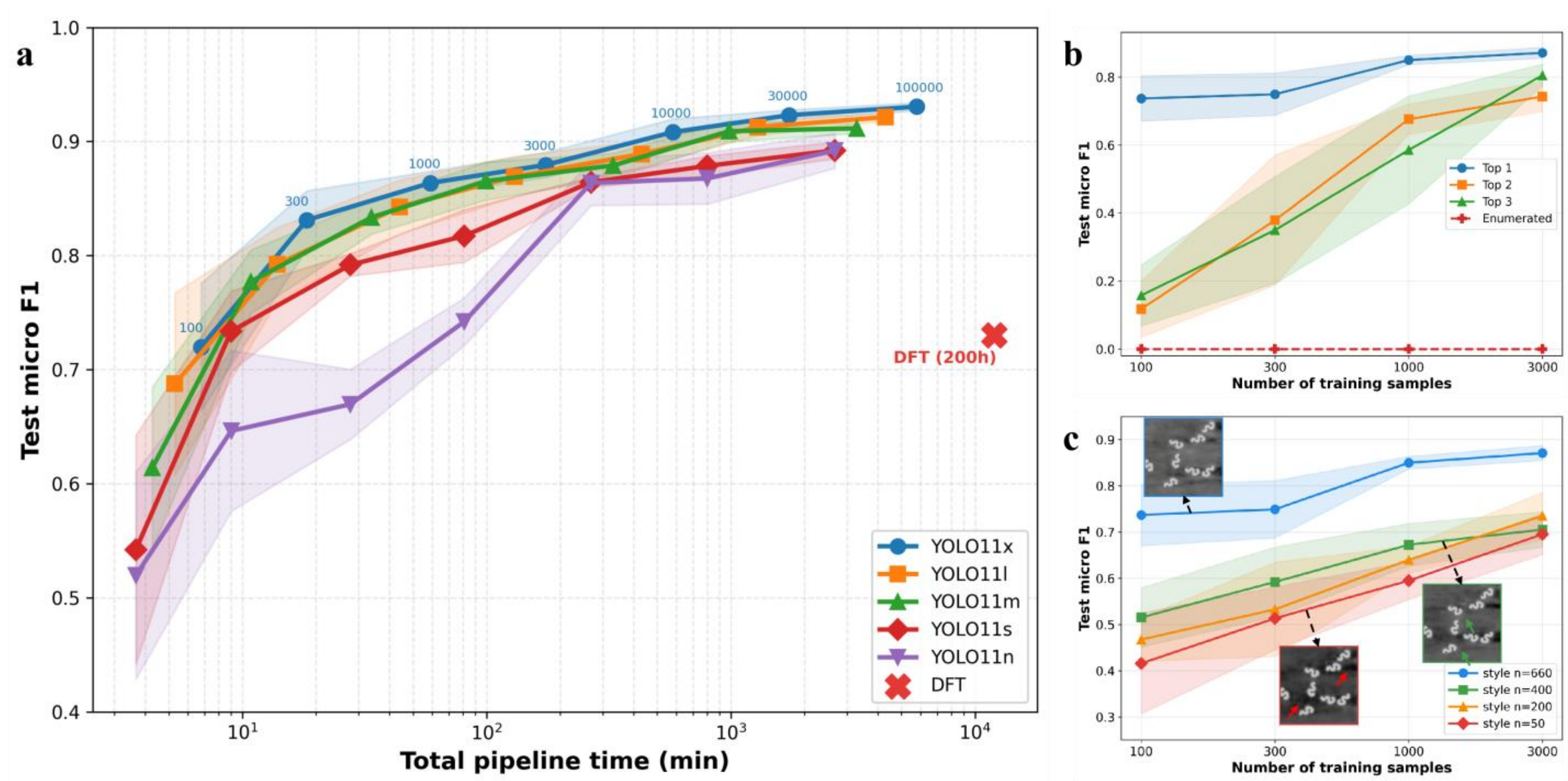

**Figure 7.** Scaling behavior of **NanoSim**. (a) synthetic-data volume and YOLO11 model size, (b) Top-k reaction-site selection, and (c) STM data volume for Pix2Pix training.

We presented a simulation-assisted framework for zero-shot STM analysis that integrates physics-grounded synthetic data generation with automated hierarchical interpretation. Using only precursor structures, without target-system STM images or annotations, the framework achieved accurate molecule-level recognition on real STM images and converted these predictions into system-level observables, including annealing-dependent population changes, and used these observables to infer reaction pathways and other mechanistic insights. By factorizing reaction constraints, surface behavior, and imaging appearance into independently learnable modules, it embeds the relevant physical and chemical priors into synthetic-data generation, eliminating the need for target-system-specific supervision and thereby enabling zero-shot generalization. This modularity further allows heterogeneous reaction, surface, and imaging data from other systems to improve the corresponding components independently as data and models evolve. Together, these capabilities provide a scalable route for quantitative analysis of new on-surface reaction systems.

## Supplementary Information

Experimental and computational methods; model training and evaluation details; ablation studies; additional molecular assignments, spatial analyses, annealing results, limitations, and transfer studies.

## Data availability

The data supporting the findings of this study are provided in the Supplementary Information, except for the raw STM images, which are available from the corresponding authors upon reasonable request.

## Code availability

The custom code used in this study is available to editors and reviewers via a private repository link during peer review and will be made publicly available after acceptance.

**Funding**

This work was supported by the National Natural Science Foundation of China (grant no. 22302076).

**Author contributions**

Y.H. conceived the main research idea and carried out the machine-learning work. Z.W. performed the scanning tunneling microscopy experiments and acquired the experimental images. Y.B., T.L. and H.B. contributed to the overall experimental design, discussed and interpreted the results, and critically revised the manuscript. All authors reviewed and approved the final manuscript.

**Competing interests**

The authors declare no competing interests.

# Supplementary Information

## Physics-Grounded Simulation Enables Zero-Shot Molecular and Reaction Analysis in Scanning Tunneling Microscopy

*Yuanqin He,[1] Zhiyu Wang,[2] Yang Bao,[3] Tao Lin,[2]* and Hai Bi[1]**

[1] Jihua Laboratory, Foshan, P. R. China

[2] College of New Materials and New Energies, Shenzhen Technology University, Shenzhen, P. R. China

[3] State Key Laboratory of Luminescence Science and Technology, Changchun Institute of Optics, Fine Mechanics and Physics, Chinese Academy of Sciences, Changchun, P. R. China

* e-mails: lintao@sztu.edu.cn; bihai@jihualab.ac.cn

# S1 Supplementary Methods

## S1.1 Experimental and Computational Data Preparation

### S1.1.1 STM Experiments

All sample preparation and STM measurements were performed in a commercial ultrahigh-vacuum system from Physcryo (Suzhou) with a base pressure below $5 \times 10^{-10}$ mbar. The Cu(111) single-crystal surface was prepared by repeated cycles of argon ion sputtering and annealing at about 700 K. 2-Bromo-1,10-phenanthroline (BP) molecules (purity: 99.4%) were purchased from Alfa Aesar and evaporated by organic molecular beam epitaxy from a quartz glass crucible held at 320 K. STM measurements were performed at 78 K, and the acquired STM images were processed using WSxM software[1].

### S1.1.2 DFT Calculations

Density functional theory calculations were performed using the Vienna Ab initio Simulation Package (VASP) [2] with the Perdew-Burke-Ernzerhof (PBE) exchange-correlation functional.[3] The electron-ion interaction was described using the projector augmented-wave (PAW) method.[4] DFT-D3 was used to account for van der Waals (vdW) interactions.[5]

To simulate molecules adsorbed on the surface, a two-layer Cu(111) slab was constructed with the bottom layer fixed. A vacuum region larger than 15 Å was applied along the z direction to prevent interactions between periodic images. The plane-wave energy cutoff was set to 450 eV. The systems were fully relaxed until the residual force on each atom was less than 0.02 eV/Å. Structural visualizations were generated using VESTA,[6] and simulated STM images were produced using the p4vasp software package (https://vasp.at/info/resource/p4vasp/).

## S1.2 NanoSim Training Data Generation

**NanoSim** organizes synthetic STM data generation into three abstraction layers that progressively constrain the product, scene, and image spaces. The Physical layer converts a precursor structure into a reaction-guided candidate product set. The Behavior layer instantiates these products as molecular identities, positions, and orientations within a surface scene. The Rendering layer converts each instantiated scene into an STM-like image while retaining automatically generated, geometrically registered annotations. Thus, the three layers transform a chemical precursor into paired synthetic images and labels while avoiding exhaustive enumeration of chemical products, surface arrangements, and image appearances.

### S1.2.1 Physical Layer: Reaction-Guided Product-Space Construction

The Physical layer constructs a reaction-guided candidate product space from the precursor structure. **NanoSim** uses a reaction-site prediction model as a chemical prior to focus product generation on highly ranked candidate transformations and limit irrelevant or redundant recognition classes. Starting from a precursor molecule $p$, each reaction round predicts reactive sites and graph edits, generates candidate products from the selected edits, and propagates the retained products to the next round. The union of all valid products accumulated to the predefined reaction depth defines the target product set. Unless otherwise specified, the Top-1 predicted reactive-site choice is retained and a reaction depth of 3 rounds is used for the BP system.

Reaction-site prediction is performed with RexGen, a template-free graph-based reaction prediction method.[7,8] DirectCoreFinder uses a Weisfeiler-Lehman network with attention, a hidden dimension of 300, and three message-passing layers. Atom-mapped reactants are supplied as dot-separated SMILES strings. RexGen returns up to 80 confidence-ranked bond edits encoded as atom1-atom2-bond_order records; these edits seed candidate-product construction. In the standard workflow, only the highest-ranked reactive-site choice is propagated, whereas the Top-k ablation retains additional ranked choices to quantify the effect of expanding the product space. Notably, the current reaction prediction model does not explicitly account for substrate effects, as established substrate-aware reaction prediction methods for on-surface reactions remain limited.

Candidate products are subjected to surface-specific diversification and filtering before entering the target product set. A surface-planarity approximation is first applied because the downstream STM representation depends primarily on projected molecular morphology. Products with invalid planar embeddings, including severe bond crossings or poor atom placement, are removed. Relevant mirrored surface arrangements are then generated to account for adsorption-induced mirror configurations, and redundant molecular representations are eliminated by structural deduplication. The resulting output of the Physical layer is a compact, surface-confined set of molecular product classes.

Supplementary Figure 1 shows how predicted reactive sites are enumerated, deduplicated, and propagated between rounds. The retained structures define the downstream recognition classes and the class-wise sampling pool used for synthetic-scene generation.

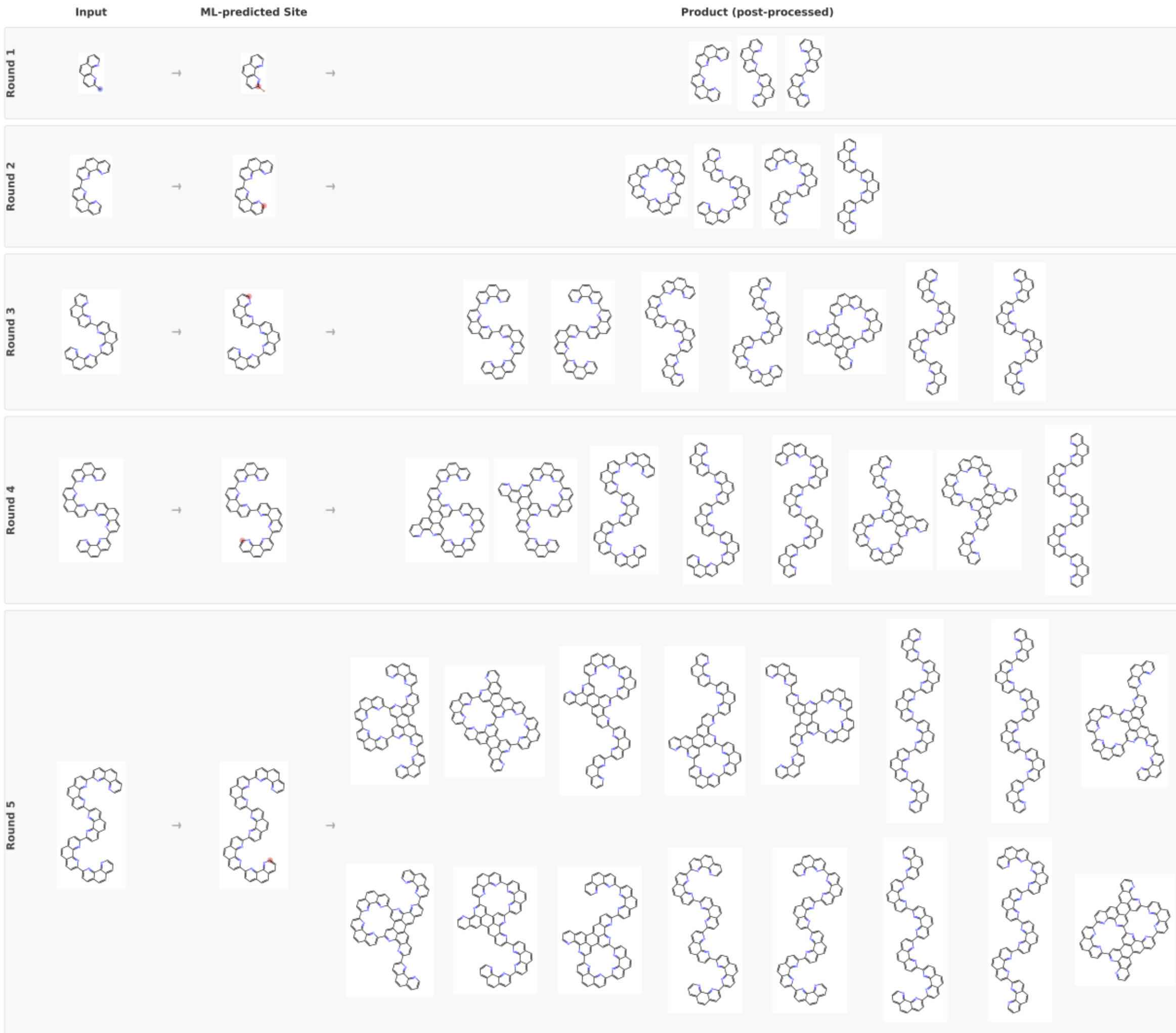

**Supplementary Figure 1. Three-round reaction-guided product-space construction for BP.** Columns show candidate molecules, predicted reactive sites, enumerated products, and mirror-deduplicated retained products; retained products advance to the next round.

### S1.2.2 Behavior Layer: Surface-Scene Construction

The Behavior layer converts the candidate product classes generated by the Physical layer into instantiated molecular scenes. It provides a compact and controllable abstraction of surface organization. A scene is parameterized by molecular number density $d_s$, minimum edge-to-edge distance $d_m$, a dimensionless molecular-sampling parameter $s_p$, and the angular sampling step $\Delta\theta$. For a square field of view with side length L, the target number of molecules is $N_{\mathrm{mol}}$ = floor($L^2$ $d_s$), where $d_s$ is expressed in molecules nm$^{-2}$. Thus, L = 20 nm and $d_s$ = 0.03 molecules nm$^{-2}$ give ~12 target molecules.

The dimensionless molecular-sampling parameter $s_p$ controls the relative sampling probability of candidate molecular classes according to their molecular complexity. Each candidate class $i$ is assigned a molecular level $\ell_i$, defined by its reaction-product group, with higher levels corresponding to structurally more complex products. For $s_p > 0$, the relative sampling weight of class $i$ is defined as:

$$w_i = {s_p}^{\ell_i}$$

The corresponding class-wise sampling probability is obtained by normalization over all candidate classes:

$$P_i = \frac{{s_p}^{\ell_i}}{\sum_j {s_p}^{\ell_j}}$$

Thus, $0 < s_p < 1$ preferentially samples lower-level, structurally simpler products, $s_p = 1$ gives equal sampling probability to every candidate class, and $s_p > 1$ preferentially samples higher-level, more complex products. In the implementation, $s_p = 0$ is treated as a special unweighted setting in which all candidate classes are sampled uniformly. Because all classes within the same molecular level have the same weight, $s_p$ controls the relative preference among individual candidate classes according to molecular complexity.

For each sampled molecule, position is drawn uniformly within the available field after accounting for its vdW bounding box, and the in-plane angle is drawn from the configured orientation range (0°-360° in the unrestricted setting). The angular sampling step $\Delta\theta$ controls the discretization of this range, such that candidate orientations are sampled at increments of $\Delta\theta$. The effects of $\Delta\theta$ = 1°, 3°, 10°, and 30° were evaluated in the ablation study. At the 20 nm, 512-pixel reference setting, the conversion factor is 25.6 pixels $nm^{-1}$. Candidate molecular masks are first tested for direct overlap. Up to 100 local placement attempts are made per molecule; a molecule is skipped if no valid placement is found.

The Behavior layer outputs an instantiated scene containing the class identity, position, orientation, projected footprint, and atomic coordinates of every accepted molecular instance, as illustrated by the molecular-placement stage in Supplementary Fig. 2. This parameterization enables controlled sampling of scene-level spatial statistics and systematic evaluation of how density, separation, composition, and orientation affect transfer from synthetic data to experimental STM images.

### S1.2.3 Rendering Layer: Geometry-Based Rendering and Style Alignment

The Rendering layer converts the instantiated molecular configurations produced by the Behavior layer into STM-like observations. Because full electronic-structure STM calculations are difficult to scale to large product and scene spaces, **NanoSim** first generates geometry-based vdW height maps and then applies a Pix2Pix-based style alignment step to translate them into experimental-like STM images.[9] The vdW representation preserves the projected molecular footprint and relative protrusions required for structural recognition, whereas style alignment adapts the contrast and texture statistics toward those of experimental STM images. Together, these sequential steps balance structural fidelity, appearance realism, and computational scalability (Supplementary Fig. 2).

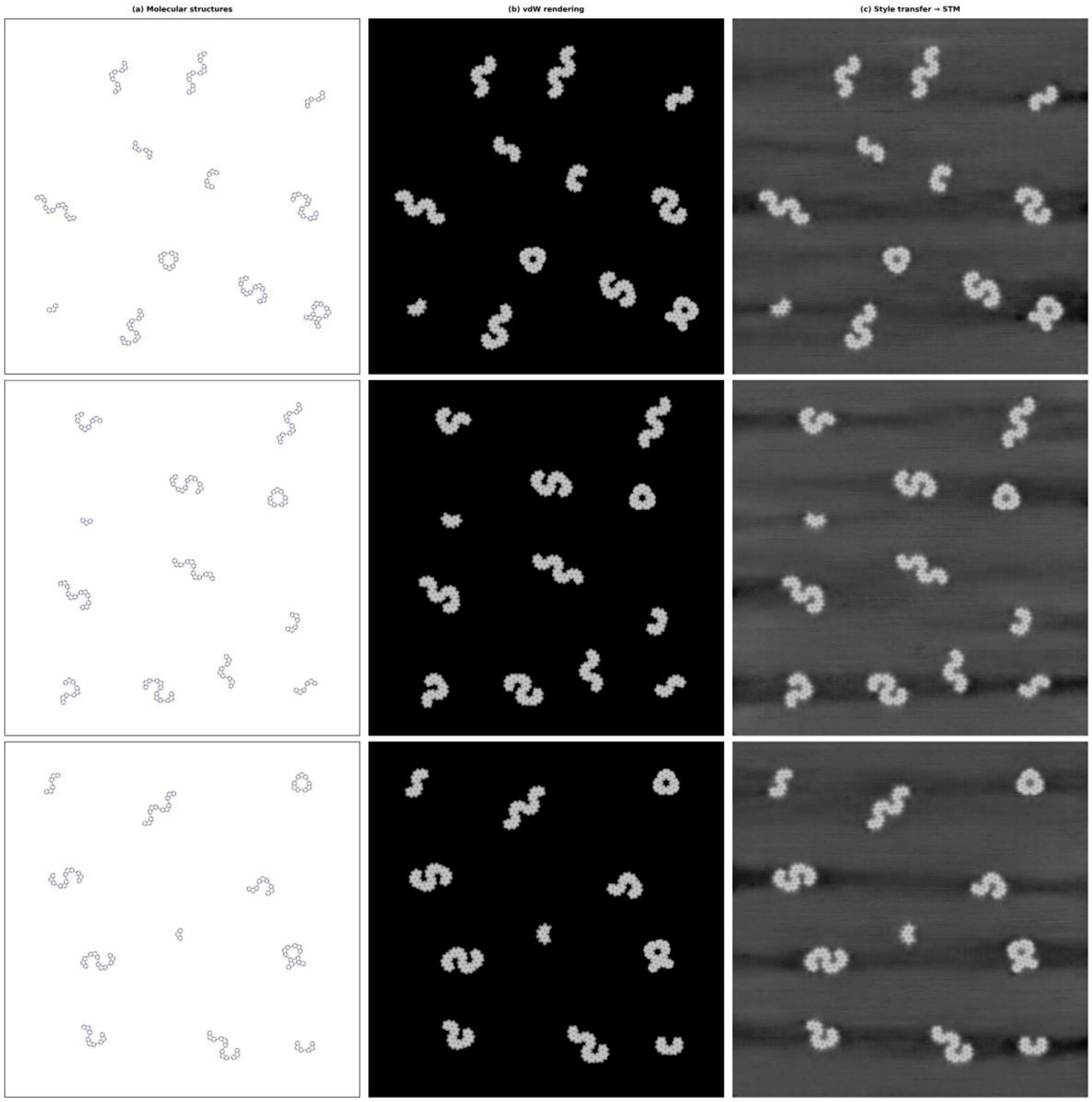

**Supplementary Figure 2. Synthetic STM image-generation pipeline.** Columns show molecular placement, vdW-sphere rendering, and Pix2Pix-based style alignment plus augmentation; rows show independent scenes. Image scale: 20×20 nm$^2$.

In the first stage, each atom is represented as a van der Waals sphere with an element-specific radius $r_i$ and center ($x_i$, $y_i$, $z_i$) in a surface-fixed coordinate system, where z is normal to the substrate after applying the adsorption pose defined by the Behavior layer. On a regular lateral grid (x, y), atom i contributes the upper spherical cap

$$h_i(x, y) = z_i + \sqrt{r_i^2 - (x - x_i)^2 - (y - y_i)^2},$$

which is defined only where $(x - x_i)^2 + (y - y_i)^2 \leq r_i^2$. The resulting approximate height image is taken as the envelope of all atomic caps,

$$h(x, y) = \max_i h_i(x, y),$$

followed by referencing to the bare-surface plane and optional clipping/normalization to match the experimental height range.

Style alignment is performed with Pix2Pix using a U-Net 256 generator and a PatchGAN discriminator to translate registered grayscale geometry-based inputs into experimental-like STM images (Supplementary Fig. 3).[9,10] Approximately 3000 candidate STM images from non-target molecular systems were initially assembled. Images that did not meet the predefined requirements for image dimensions, spatial scale, and image quality were excluded, leaving approximately 660 images. Adaptive thresholding was applied to each retained image to generate a spatially registered threshold-processed input, while the corresponding grayscale STM image served as the target. The model was trained for 100 epochs on these 660 paired images, resized to 512 × 512 pixels. None of the molecular species evaluated here, including BP and those in the Supplementary Information, were included in the Pix2Pix training set, thereby ensuring zero-shot evaluation.

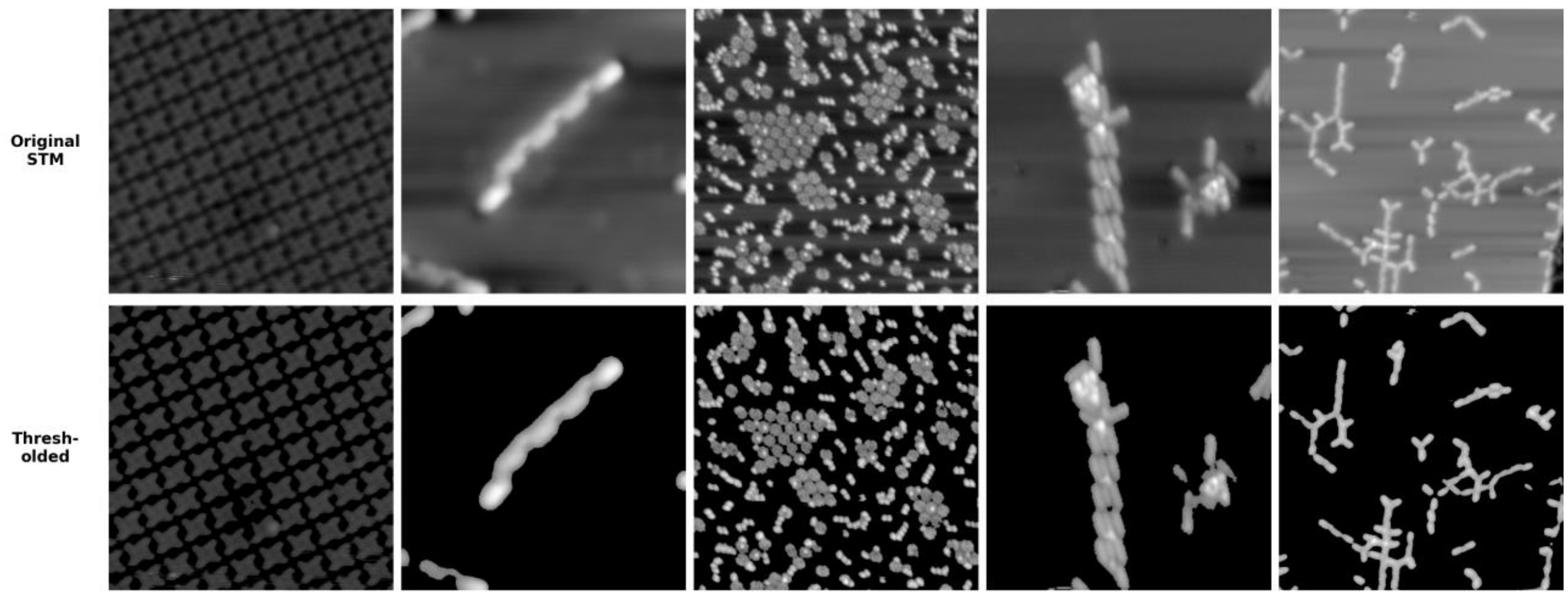

**Supplementary Figure 3. Representative paired inputs for style alignment.** Experimental STM targets (B) are shown in the top row, and the corresponding registered threshold-processed inputs (A) are shown in the bottom row.

The Rendering layer outputs an STM-like synthetic image together with geometrically registered annotations derived from the known molecular identities, footprints, poses, and atomic coordinates. Bounding boxes, oriented bounding boxes, instance masks, and keypoints can therefore be generated without manual annotation and transformed together with the image during subsequent augmentation.

Hand-crafted transformations that emulate tip convolution, background slope, impurities, and experimental noise can improve the realism of synthetic STM images and have been used in previous AI-assisted STM analyses.[11] In **NanoSim**, such operations can serve as auxiliary augmentations, whereas the required Pix2Pix style alignment step provides a data-driven route for learning the combined appearance signatures associated with the instrument, tip state, and substrate.

### S1.2.4 Synthetic Label Generation

Because each synthetic STM image is generated from known molecular identities, poses, and atomic coordinates, labels for all recognition tasks can be produced automatically without manual annotation. For object detection, the projected molecular footprint is enclosed by the smallest axis-aligned rectangle to define the bounding box. For oriented bounding box (OBB) detection, a class-specific molecular box is first defined in the molecular coordinate frame and then transformed according to the sampled in-plane position and rotation. For instance segmentation, the mask is obtained from the molecular region occupied by each simulated instance in the rendered image. For keypoint localization, the image-plane centers of five- and six-membered rings are used as representative structural landmarks. When geometric transformations are applied during synthetic-data generation, all labels are transformed together with the image to preserve image-annotation registration.

From a single synthetic scene, **NanoSim** automatically derives geometrically registered BBox, OBB, keypoint, and instance-mask annotations, allowing the same scene to supervise multiple perception tasks (Supplementary Fig. 4).

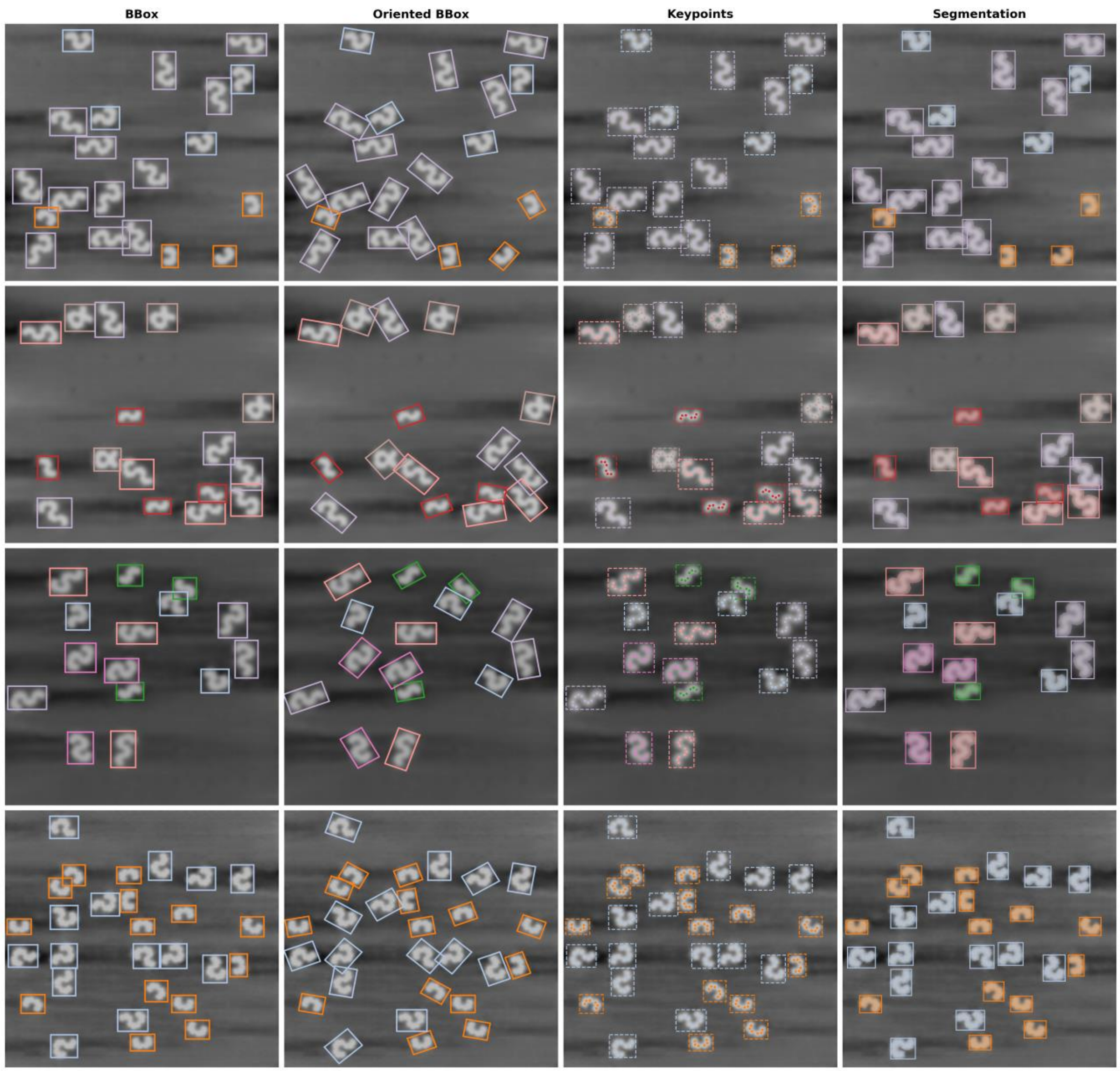

**Supplementary Figure 4. Geometrically registered annotations generated from synthetic scenes.** Columns show BBox, OBB, keypoint, and instance-mask labels; rows show independent synthetic scenes. Class colors are consistent across formats. Image scale: 20×20 $nm^2$.

## S1.3 Model Training

YOLO11 provides task-specific variants for object detection, instance segmentation, keypoint localization, and oriented bounding-box detection, with five model sizes-n, s, m, l, and x-of increasing capacity.[12] All models in this study were trained on the corresponding synthetic datasets for 100 epochs with early stopping disabled; the complete training and evaluation settings are summarized in Supplementary Table 1.

Because the synthetic training and validation sets are sampled from the same simulated domain, conventional synthetic-validation metrics can remain high even when transfer to experimental STM images is suboptimal. Labels from the real target-system images were not used during model training under our setting. We therefore applied the same fixed 100-epoch schedule to all reported dataset sizes and model capacities, with early stopping disabled.

To assess whether the fixed training schedule was sufficient, we retrospectively examined representative training trajectories across several synthetic dataset sizes and model capacities (Supplementary Fig. 5). Real-image micro-F1 generally reached a plateau by 100 epochs, supporting the adequacy of the fixed schedule for the reported configurations. This retrospective comparison was not used to select the reported checkpoints.

To identify a metric that could monitor convergence without using labeled real images during training, we also compared several statistics recorded on the synthetic validation set. Among the evaluated metrics, validation confidence showed the clearest correspondence with model convergence, approaching a plateau near the epoch at which real-image micro-F1 stabilized (Supplementary Fig. 5). Validation confidence may therefore serve as a practical early-stopping signal. Terminating training once this metric has stabilized could further reduce the total training time relative to the fixed 100-epoch schedule.

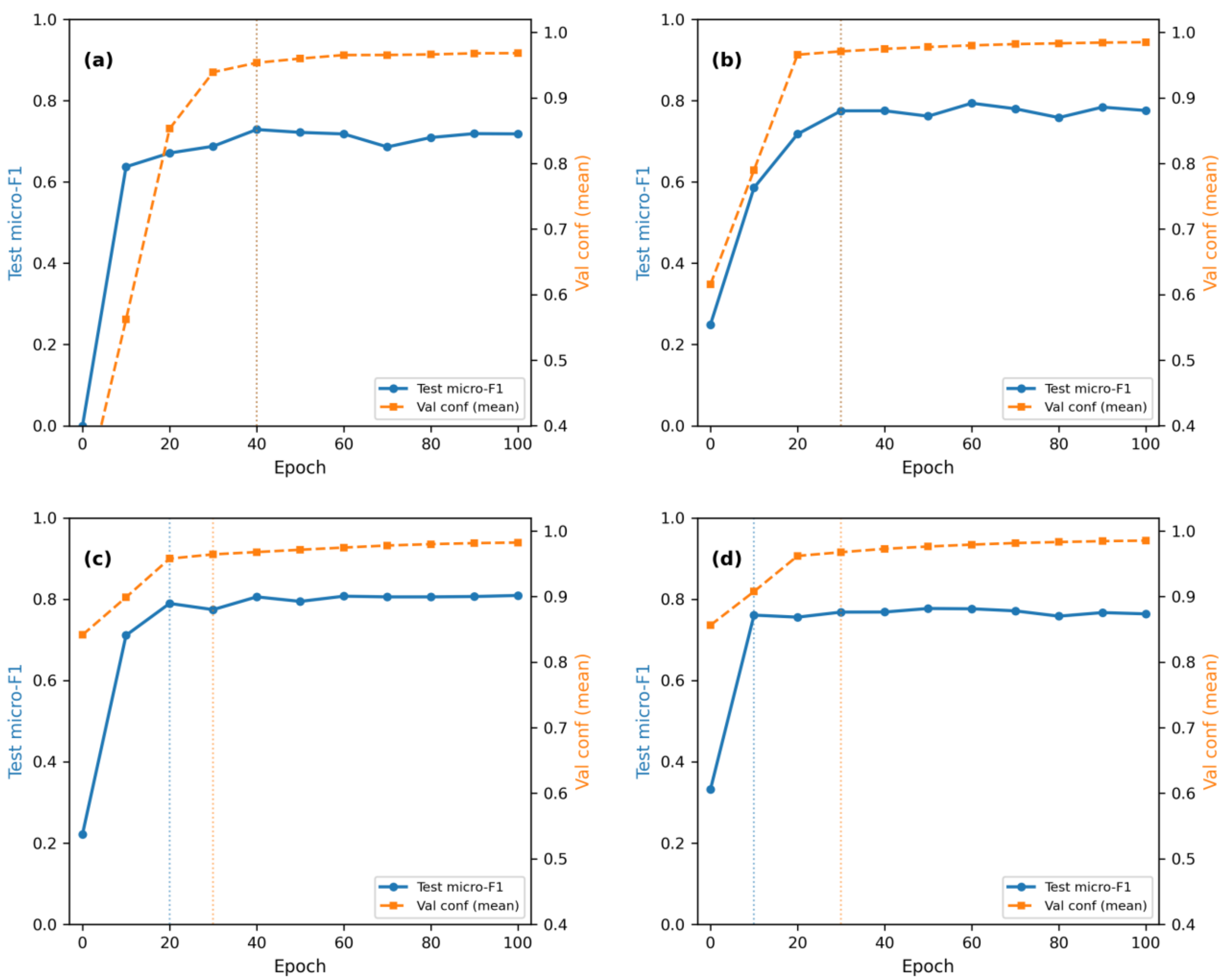

**Supplementary Figure 5. Validation-confidence and real-image test trajectories.** Trajectories are shown for models trained with increasing synthetic dataset sizes.

**Supplementary Table 1. Core training and evaluation settings.**

| **Category** | **Parameter** | **Value** |
|---|---|---|
| Training | Epochs | 100 |
| | Batch size | 8 |
| | Input image size | 512 × 512 pixels |
| | Optimizer and schedule | AdamW; cosine schedule |
| | Initial learning rate | 0.01 |
| | Weight decay | 0.0005 |
| | Augmentation | Disabled: Mosaic, MixUp, flips, translation, and rotation |
| Evaluation | Confidence / IoU threshold | 0.5 / 0.5 |
| | NMS | Class-agnostic |

## S1.4 Evaluation and Scale-Adaptive Inference

### S1.4.1 Evaluation Metrics

Micro-F1 was used as the primary aggregate metric for recognition in real STM images. It is calculated by pooling the true-positive, false-positive, and false-negative detections from all classes and images, computing one overall precision and recall from these pooled counts, and taking their harmonic mean. Each annotated molecular instance therefore contributes to the final score according to the detection outcome, rather than each molecular class receiving the same weight.

This instance-level aggregation is important because **NanoSim** generates a relatively broad reaction-guided candidate class space, whereas an experimental STM dataset normally contains only an unknown subset of these products. The generated or predicted class space and the classes actually present in the test images can therefore differ substantially. Macro-F1 first calculates F1 for each class and then gives all classes equal weight; in the present setting, classes absent from the test set do not provide a meaningful class-wise recall or F1, while scores for rare classes can vary strongly because they contain very few instances. mAP also evaluates performance separately for each class from ranked predictions before averaging across classes, so it is similarly sensitive to which classes are present and to the limited support of rare classes. In addition, mAP summarizes detection quality over a range of confidence settings, whereas the downstream population, orientation, and distance analyses use detections at a fixed operating point. Micro-F1 therefore provides a more direct and stable summary of the final detections used in this work: objects from observed classes contribute according to their frequency, and predictions assigned to classes absent from the test set are still counted as false positives.

For the reported evaluation, predictions with confidence scores of at least 0.5 were matched to same-class ground-truth objects using an Intersection over Union (IoU) threshold of 0.5. Each prediction and ground-truth object could be matched only once; unmatched predictions and annotations were counted as false positives and false negatives, respectively.

For each training configuration, model training was repeated using five independently generated synthetic-data seeds, and all resulting models were evaluated on the same fixed real-image test set. Reported means and standard deviations therefore quantify variability associated with synthetic-data generation and model training, rather than uncertainty arising from experimental-image sampling.

### S1.4.2 Test-Set Composition

The real STM test set comprised 10 images randomly selected from STM images of sufficient quality, yielding 515 molecular instances. One researcher manually assigned the bounding-box and oriented-bounding-box labels, and a second researcher reviewed all annotations. Disagreements were resolved by consensus. Keypoint and segmentation labels were not evaluated quantitatively because they are difficult to annotate unambiguously in experimental STM images.

Supplementary Figure 6 summarizes the test-set composition by molecular class and oligomer order. The class distribution is strongly imbalanced, with m4 (n = 236) and m1 (n = 138) comprising most annotations. By oligomer order, the dataset contains 47 monomers, 149 dimers, 250 trimers, 53 tetramers, 13 pentamers, and 3 hexamers.

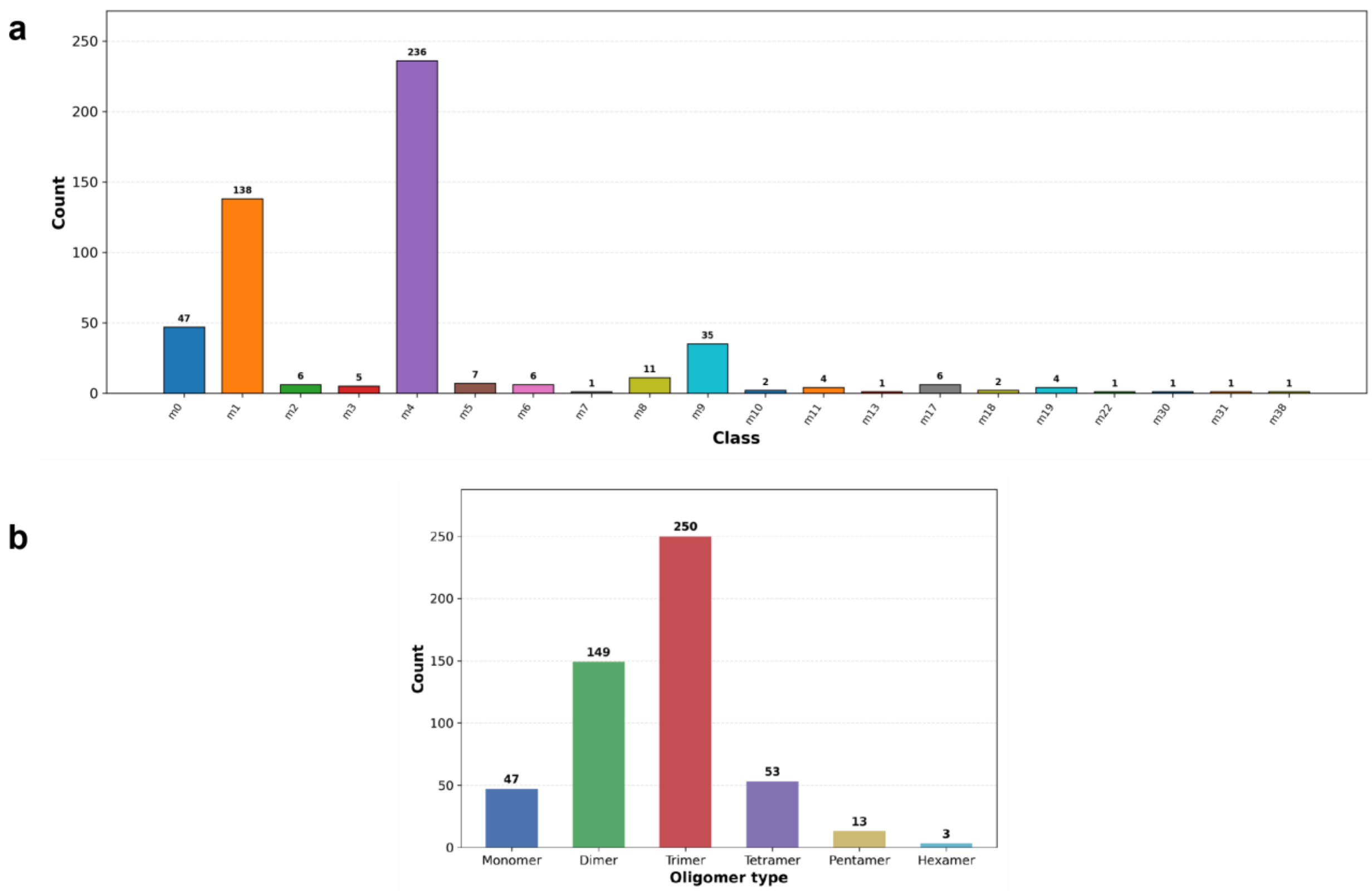

**Supplementary Figure 6. Composition of the manually annotated real STM test set.** (a) Molecular-instance counts by class. (b) Counts by oligomer order.

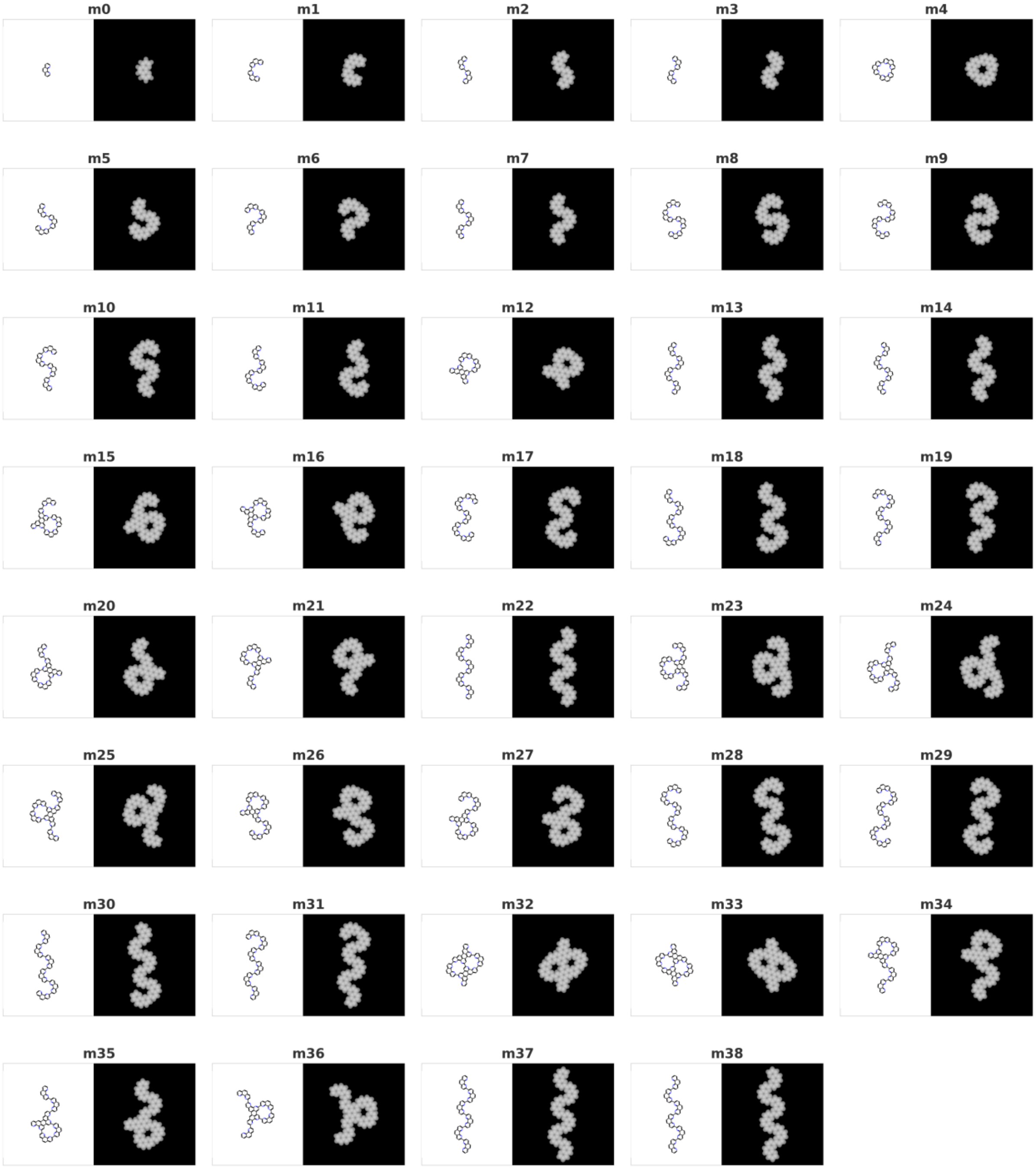

**Supplementary Figure 7. Molecular identifiers and corresponding target structures.** Structures define the classes used for labeling, training, and evaluation.

### S1.4.3 Scale-Adaptive Inference and Coordinate Transformation

A key advantage of the framework is its ability to process STM images with different physical fields of view and pixel resolutions. This capability is important for practical STM analysis, where datasets are often collected using different scan sizes, sampling resolutions, and imaging conditions. We used a model trained with 20 nm × 20 nm, 512 × 512 pixel images as the reference setting and applied it to two representative field-of-view conditions (10 nm × 10 nm and 40 nm × 40 nm), as well as to images with different pixel resolutions (Supplementary Fig. 8). The preprocessing and coordinate-transformation procedure is summarized in Supplementary Fig. 9, and the quantitative size-resolution ablation is reported in Supplementary Table 8.

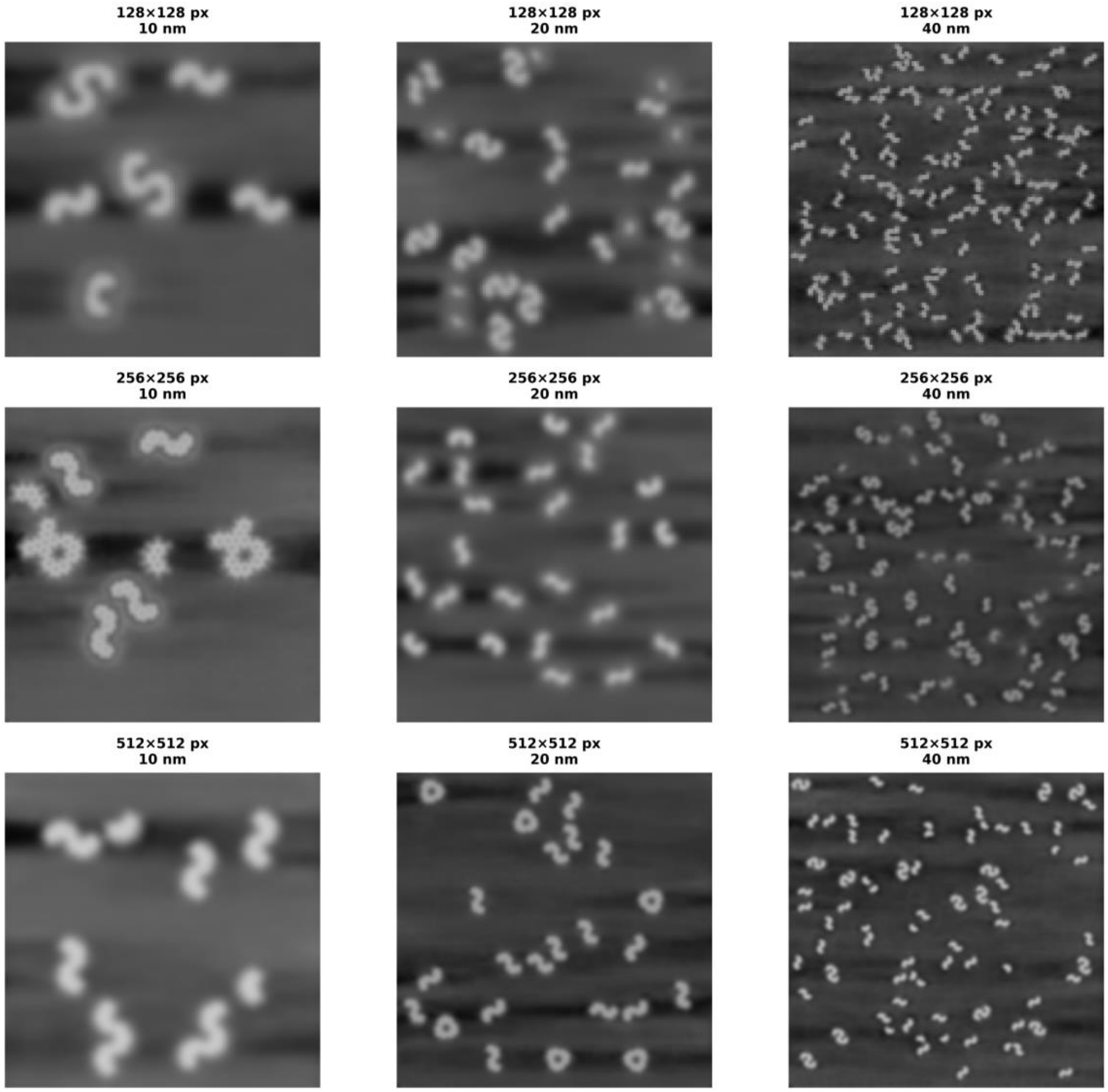

**Supplementary Figure 8. Synthetic STM images at different scan sizes and pixel resolutions.** Rows and columns compare 10, 20, and 40 nm fields of view with 128, 256, and 512 pixel representations.

This cross-scale application is achieved through preprocessing and coordinate transformation. Each input image is converted to the 20 nm × 20 nm, 512 × 512 pixel reference format for inference, and the predicted coordinates are then mapped back to the original image. For smaller fields of view, the image is padded and resized before prediction, then cropped and rescaled to the original size. For images with the same physical field of view but different pixel resolutions, only resizing to and from the reference resolution is required. For larger fields of view, the image is split into 20 nm × 20 nm tiles for independent prediction, and the tile-level results are subsequently merged. Differences in physical field of view and pixel sampling are therefore accommodated without retraining the model.

This figure describes the coordinate transformations that enable one reference model to process experimental images with different physical fields of view and pixel resolutions. The preprocessing changes image geometry for inference and then maps detections back to the original coordinate system.

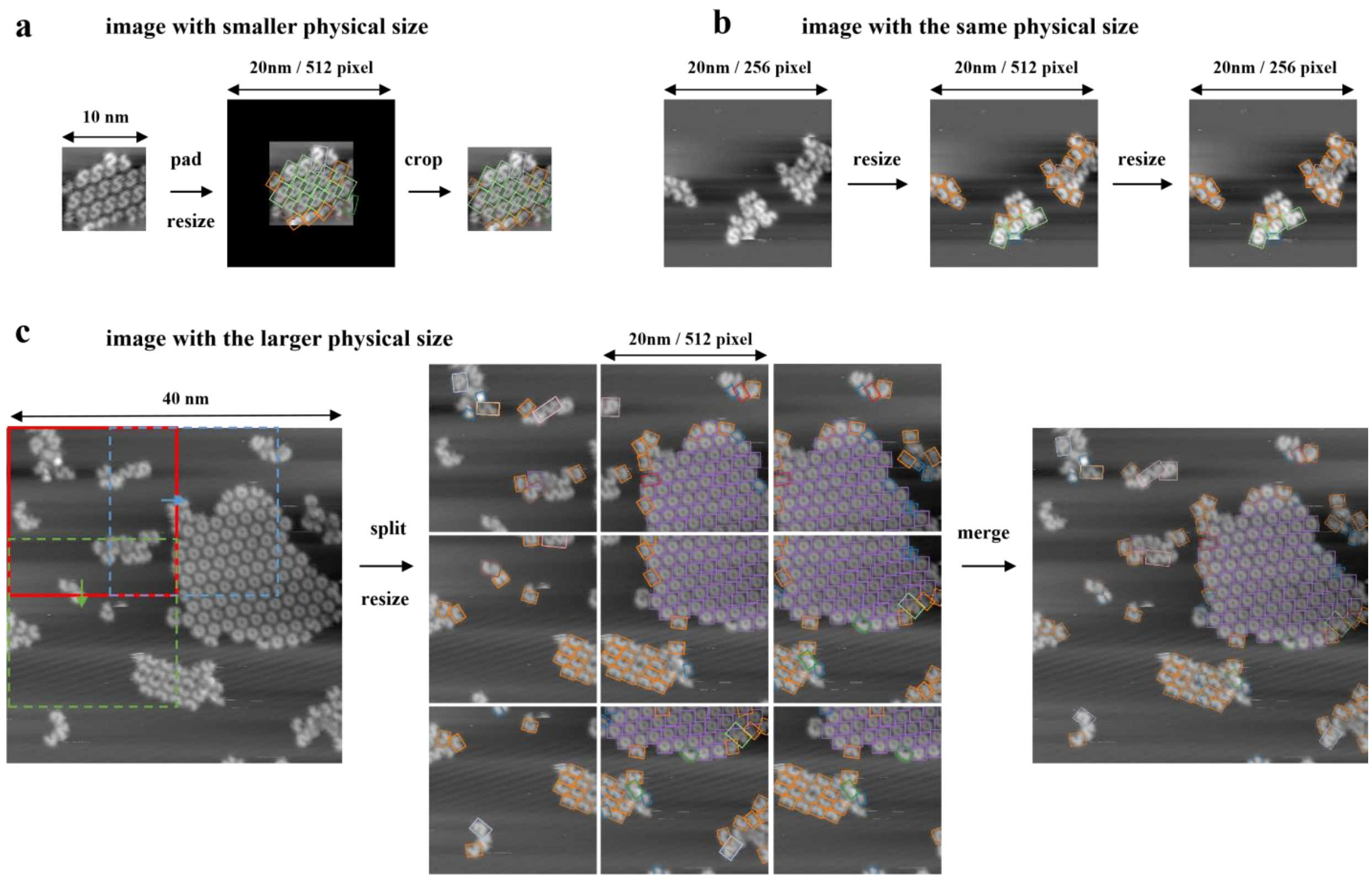

**Supplementary Figure 9. Scale-adaptive inference and coordinate transformation.** Small fields are padded and resized, different pixel resolutions are resized, and larger fields are tiled; predictions are mapped back to the original coordinates.

# S2 Supplementary Discussion

Section S2 examines four linked questions: how reaction-guided candidate-space design controls class complexity; how scene and rendering choices affect transfer to real STM images; how scale, model capacity, and computational cost trade off; and what molecular observables can be extracted from experimental STM data.

## S2.1 Reaction-Guided Candidate-Space Construction

### S2.1.1 Multi-Round Product-Space Construction and Class-Space Complexity

Reaction-guided restriction produces a substantially more learnable recognition space than exhaustive enumeration (Supplementary Fig. 1 and Supplementary Tables 2–3). In the Top-1 workflow, the 3-round, 4-round, and 5-round settings contain 15, 23, and 39 classes, respectively; Top-1, Top-2, and Top-3 selection over 3 rounds produces 15, 49, and 59 classes. Exhaustive enumeration yields more than 5000 classes and no meaningful downstream detection performance.

We explored how candidate-space depth affects the recognition of products generated in different reaction rounds by comparing YOLO11l models trained on 3,000 synthetic images derived from 2-, 3-, 4-, or 5-round product spaces. Supplementary Figure 10 shows the F1 score for each molecular class, whereas Supplementary Fig. 11 summarizes micro-F1 after grouping the products according to their generation round. The precursor m0 remained comparatively difficult to recognize, and m5-m7 also showed lower class-wise F1 scores than several other products. At the group level, the 2-round model achieved the highest performance for m0 and m1-m3, whereas the 3-round model performed best for m4-m7. Notably, the 4-round model outperformed the 3-round model for m8-m14, indicating that expanding the candidate space did not uniformly reduce recognition performance. This improvement may arise from additional supervision provided by structurally related products, which could help the model distinguish similar projected morphologies.

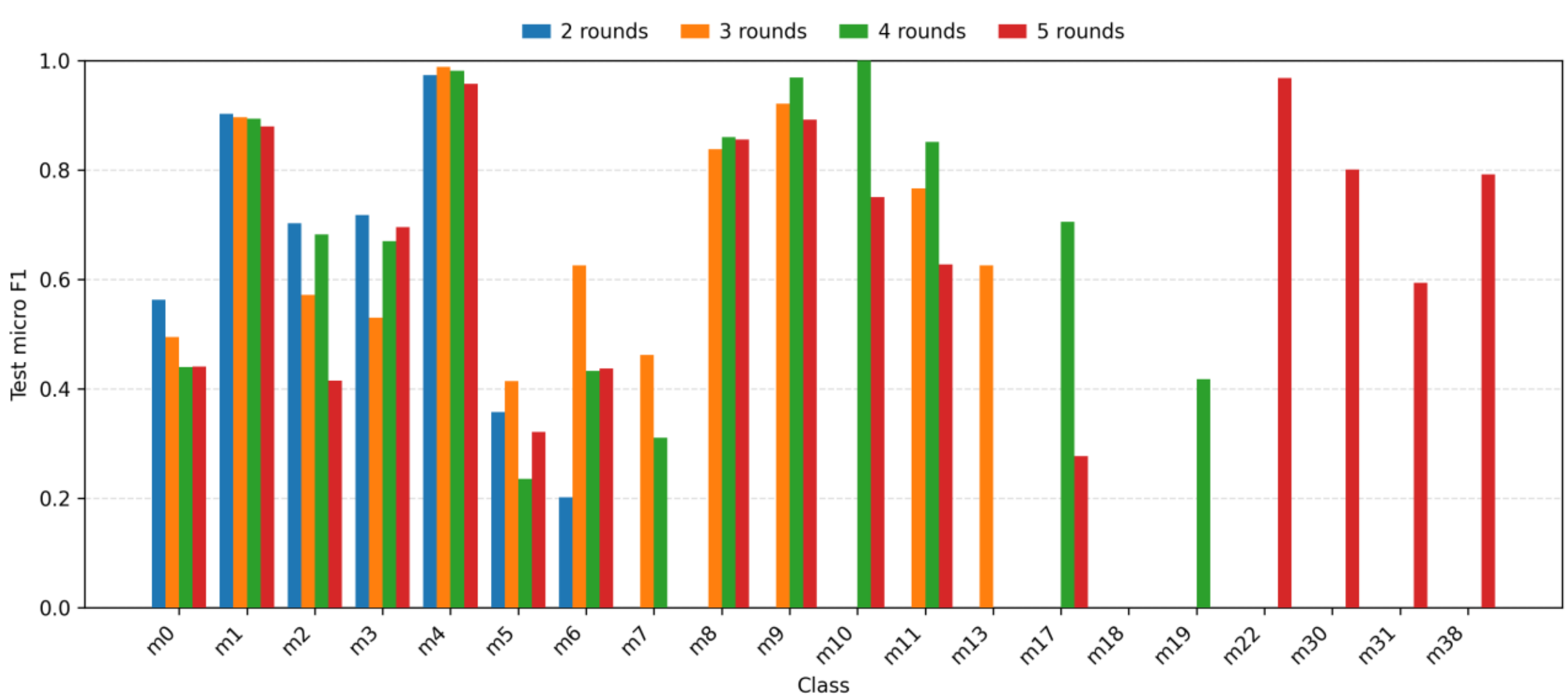

**Supplementary Figure 10. Class-wise F1 across reaction-guided product spaces of different depths.** Results are shown for YOLO11l models trained on n = 3000 synthetic images. Bars report the F1 score for each class included in the corresponding 2-, 3-, 4-, or 5-round product space; missing bars indicate classes absent from that product space.

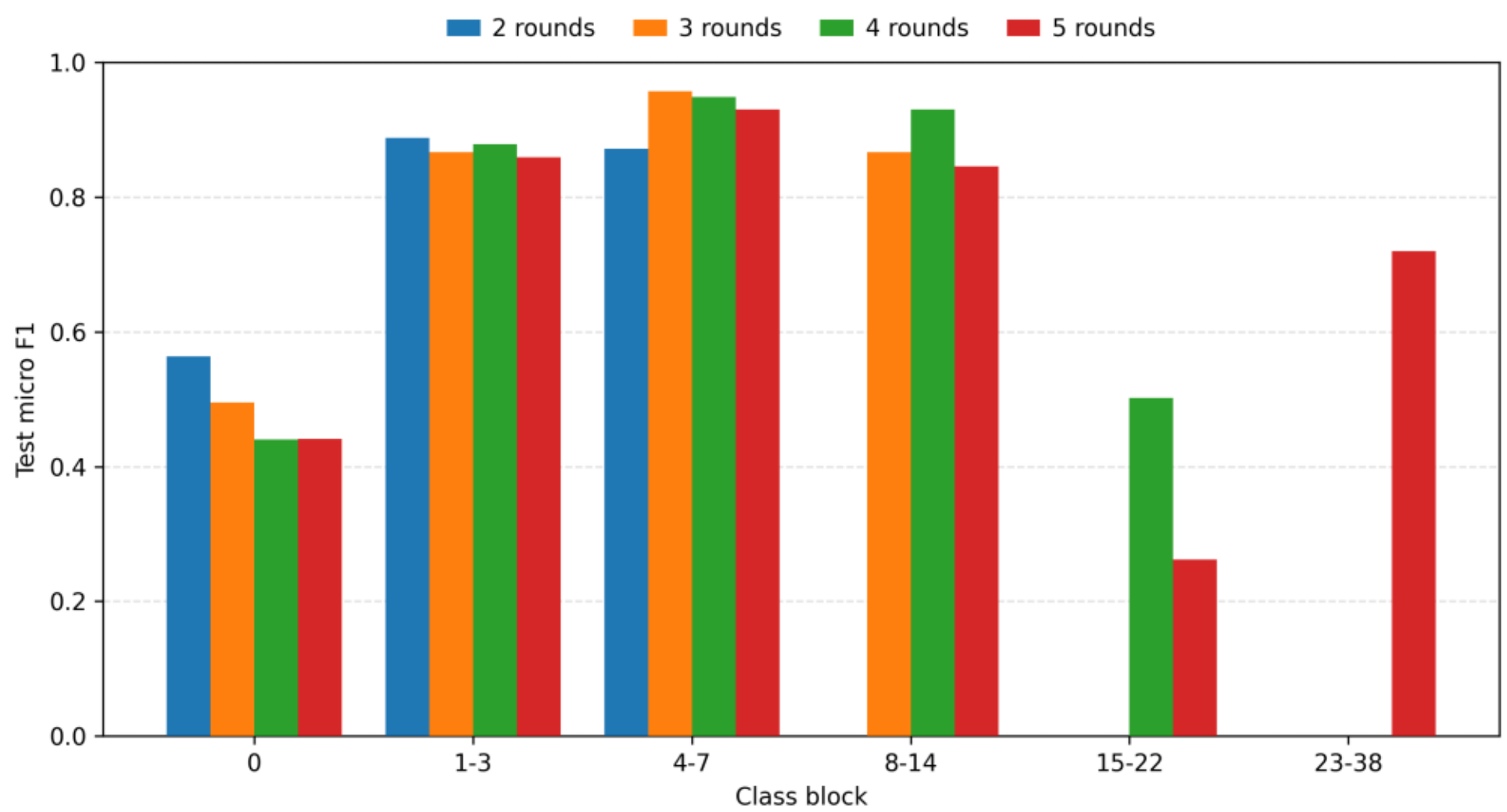

**Supplementary Figure 11. Recognition performance grouped by product-space generation round.** Results are shown for YOLO11l models trained on n = 3000 synthetic images using 2-, 3-, 4-, or 5-round candidate spaces. Bars report micro-F1 for the precursor m0 and for class blocks introduced in successive rounds (m1-m3, m4-m7, m8-m14, m15-m22, and m23-m38); missing bars indicate class blocks absent from the corresponding product space.

### S2.1.2 Top-k Reaction-Site Selection and Generated Product Spaces

Restricting product generation to the highest-ranked reaction site gives the strongest performance across the tested data volumes (Supplementary Fig. 12 and Supplementary Table 2). At n = 3000, Top-1 reaches a micro-F1 of 0.871 ± 0.016, compared with 0.743 ± 0.044 for Top-2 and 0.805 ± 0.033 for Top-3; exhaustive enumeration gives a micro-F1 of 0 at every tested data volume.

Top-2 and Top-3 introduce additional chemically plausible but experimentally rare or visually similar products, increasing class-space complexity. The performance gap narrows as the synthetic training set grows, indicating that additional data partially compensate for the enlarged candidate space.

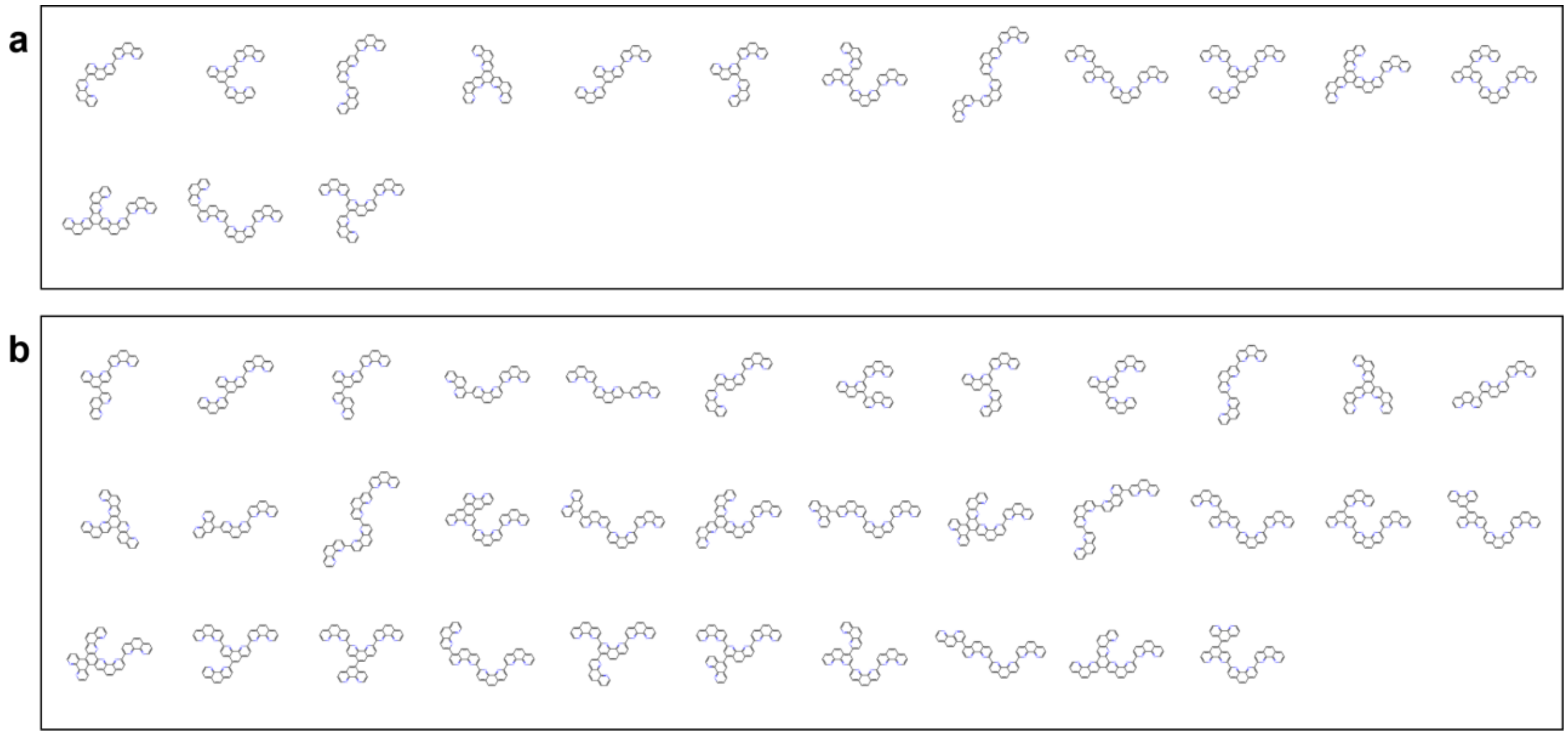

**Supplementary Figure 12. Product spaces generated using Top-2 and Top-3 reaction-site selections.** (a) Top-2 products. (b) Products added by Top-3 relative to Top-2.

**Supplementary Table 2. Effect of reaction-site selection strategy and synthetic training-set size on test micro-F1.**

| **Strategy** | **n = 100** | **n = 300** | **n = 1000** | **n = 3000** |
|---|---|---|---|---|
| Top-1 | 0.737 ± 0.067 | 0.749 ± 0.062 | 0.850 ± 0.014 | 0.871 ± 0.016 |
| Top-2 | 0.117 ± 0.083 | 0.379 ± 0.192 | 0.676 ± 0.045 | 0.743 ± 0.044 |
| Top-3 | 0.158 ± 0.090 | 0.350 ± 0.157 | 0.585 ± 0.159 | 0.805 ± 0.033 |
| Enumerated | 0.000 ± 0.000 | 0.000 ± 0.000 | 0.000 ± 0.000 | 0.000 ± 0.000 |

### S2.1.3 Candidate-Space Depth and Configuration Ablations

Recognition generally becomes more difficult as candidate-space complexity increases, although the dependence on product-space depth is not strictly monotonic (Supplementary Table 3). At n = 3000, the 2-round configuration reaches a micro-F1 of 0.915 ± 0.008, compared with 0.868 ± 0.016, 0.873 ± 0.017, and 0.794 ± 0.047 for the 3-round, 4-round, and 5-round configurations, respectively. For the 3-round configuration, increasing the training set to 10000 and 30000 images raises micro-F1 to 0.893 ± 0.012 and 0.909 ± 0.004, showing that additional data can partly offset the larger candidate space.

**Supplementary Table 3. Effect of reaction-derived candidate-space depth (2-5 rounds) on test micro-F1.**

| **Product-space depth** | **n = 100** | **n = 300** | **n = 1000** | **n = 3000** |
|---|---|---|---|---|
| 2 rounds | 0.814 ± 0.025 | 0.870 ± 0.029 | 0.889 ± 0.017 | 0.915 ± 0.008 |
| 3 rounds | 0.686 ± 0.062 | 0.736 ± 0.057 | 0.837 ± 0.013 | 0.868 ± 0.016 |
| 4 rounds | 0.692 ± 0.030 | 0.824 ± 0.038 | 0.862 ± 0.014 | 0.873 ± 0.017 |
| 5 rounds | 0.370 ± 0.179 | 0.464 ± 0.188 | 0.685 ± 0.080 | 0.794 ± 0.047 |

## S2.2 Synthetic Scene and Appearance Generation

### S2.2.1 Surface-Scene Sampling and Behavior-Abstraction Ablation

The Behavior layer controls molecular number density $d_s$, minimum intermolecular separation $d_m$, class-wise abundance, the dimensionless molecular-sampling weight $s_p$, and the step used to sample in-plane orientations $\Delta\theta$. The parameter $s_p$ controls the relative weighting used during molecular selection, with $s_p = 1.0$ corresponding to uniform sampling across individual candidate classes. Supplementary Tables 4–6report the complete parameter-ablation results as mean and standard deviation across five independent synthetic-data seeds. At n = 3000, the best molecular-sampling result is obtained with $s_p = 1.0$ (micro-F1 = 0.876 ± 0.018), while $s_p = 2.5$ gives a nearly identical 0.874 ± 0.012. Across angular sampling steps of 1°, 3°, 10°, and 30°, micro-F1 ranges only from 0.854 to 0.878 at n = 3000, with the highest value obtained using the 1° step (0.878 ± 0.007).

**Supplementary Table 4. Effect of the molecular sampling parameter $s_p$ on test micro-F1.**

| $s_p$ | **n = 100** | **n = 300** | **n = 1000** | **n = 3000** |
|---|---|---|---|---|
| 0.1 | 0.235 ± 0.133 | 0.421 ± 0.074 | 0.691 ± 0.097 | 0.766 ± 0.057 |
| 0.5 | 0.465 ± 0.197 | 0.721 ± 0.076 | 0.808 ± 0.036 | 0.857 ± 0.025 |
| 1.0 | 0.666 ± 0.130 | 0.742 ± 0.064 | 0.834 ± 0.028 | 0.876 ± 0.018 |
| 1.5 | 0.681 ± 0.085 | 0.762 ± 0.074 | 0.834 ± 0.019 | 0.867 ± 0.012 |
| 2.0 | 0.610 ± 0.135 | 0.778 ± 0.047 | 0.828 ± 0.042 | 0.860 ± 0.017 |
| 2.5 | 0.679 ± 0.057 | 0.769 ± 0.024 | 0.817 ± 0.013 | 0.874 ± 0.012 |
| 3.0 | 0.543 ± 0.111 | 0.738 ± 0.022 | 0.794 ± 0.033 | 0.851 ± 0.020 |

Supplementary Figure 13 shows how the molecular-sampling weight $s_p$ changes the class composition of the generated scenes. Values below 1 strongly concentrate sampling in the earliest classes, particularly m0, whereas $s_p = 1.0$ produces an approximately uniform distribution across m0-m14. Increasing $s_p$ above 1 shifts sampling toward later product classes. This distribution-level comparison complements the detection-performance results in Supplementary Table 4 and shows that $s_p$ controls class balance rather than the total number of sampled molecules.

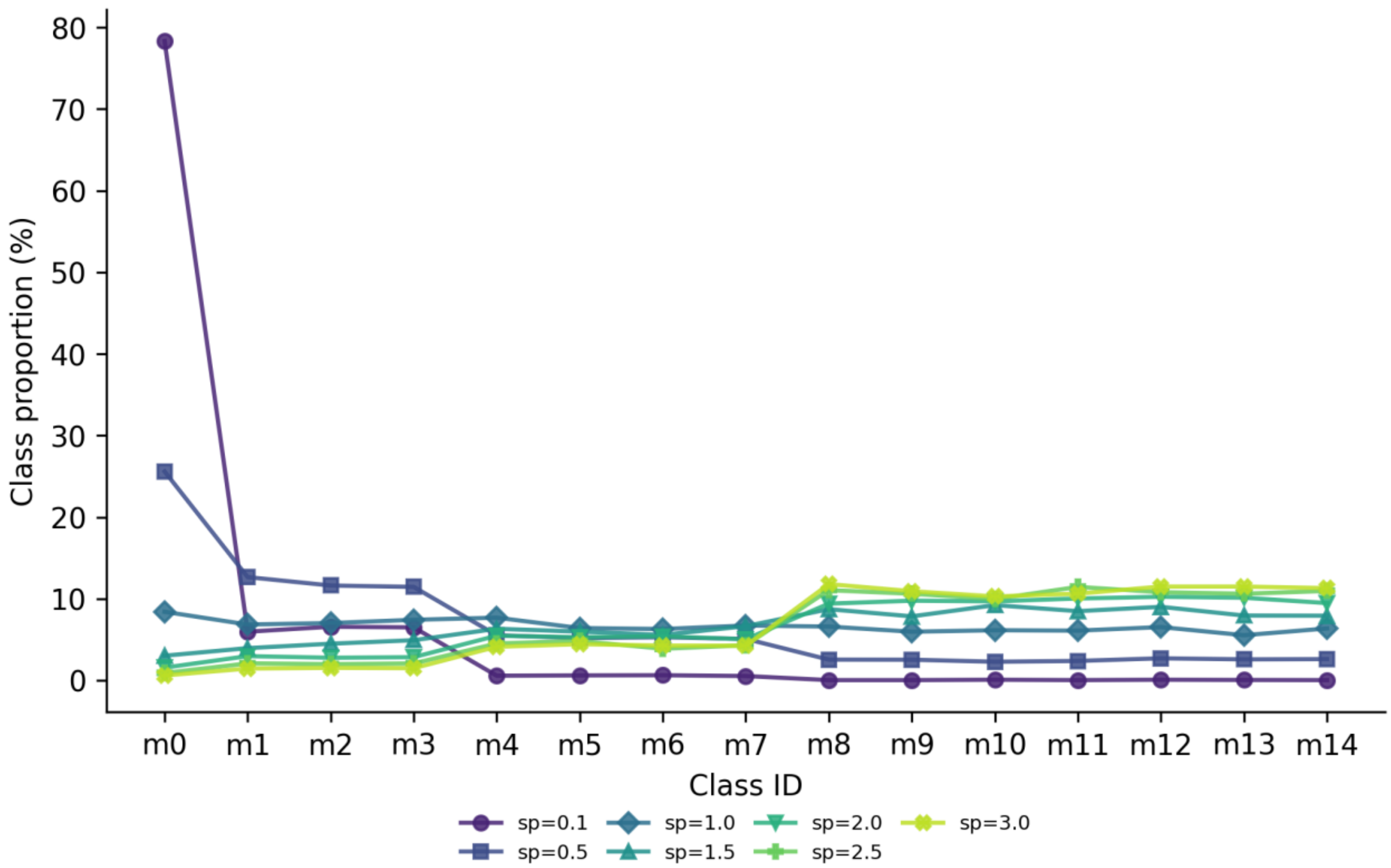

**Supplementary Figure 13. Class distributions under different molecular-sampling settings.** Class proportions for m0-m14 are shown at each tested $s_p$ value.

**Supplementary Table 5. Effect of the angular sampling step on test micro-F1.**

| angle | n = 100 | n = 300 | n = 1000 | n = 3000 |
|---|---|---|---|---|
| 1 | 0.651 ± 0.065 | 0.716 ± 0.093 | 0.839 ± 0.029 | 0.878 ± 0.007 |
| 3 | 0.631 ± 0.017 | 0.663 ± 0.060 | 0.821 ± 0.041 | 0.854 ± 0.024 |
| 10 | 0.657 ± 0.044 | 0.724 ± 0.083 | 0.846 ± 0.016 | 0.864 ± 0.013 |
| 30 | 0.664 ± 0.052 | 0.770 ± 0.038 | 0.822 ± 0.035 | 0.857 ± 0.034 |

The complete micro-F1 results for combinations of synthetic molecular density $d_s$, minimum edge-to-edge distance $d_m$, and training-set size are reported in Supplementary Table 6; their interpretation is provided in the main text.

**Supplementary Table 6. Joint effect of surface coverage $d_s$ and minimum edge-to-edge distance $d_m$ on test micro-F1.**

| $d_s$ | $d_m$ | n = 100 | n = 300 | n = 1000 | n = 3000 |
|---|---|---|---|---|---|
| 0.03 | 0.03 | 0.603 ± 0.059 | 0.706 ± 0.054 | 0.782 ± 0.014 | 0.825 ± 0.006 |
| 0.03 | 0.3 | 0.425 ± 0.181 | 0.573 ± 0.004 | 0.639 ± 0.038 | 0.696 ± 0.007 |
| 0.03 | 3.0 | 0.135 ± 0.089 | 0.260 ± 0.090 | 0.417 ± 0.237 | 0.446 ± 0.212 |
| 0.1 | 0.03 | 0.610 ± 0.064 | 0.720 ± 0.029 | 0.792 ± 0.054 | 0.860 ± 0.015 |

| $d_s$ | $d_m$ | n = 100 | n = 300 | n = 1000 | n = 3000 |
|---|---|---|---|---|---|
| 0.1 | 0.3 | 0.499 ± 0.043 | 0.542 ± 0.043 | 0.677 ± 0.008 | 0.782 ± 0.023 |
| 0.1 | 3.0 | 0.220 ± 0.094 | 0.241 ± 0.174 | 0.425 ± 0.175 | 0.561 ± 0.013 |
| 0.3 | 0.03 | 0.691 ± 0.041 | 0.812 ± 0.043 | 0.853 ± 0.031 | 0.880 ± 0.014 |
| 0.3 | 0.3 | 0.509 ± 0.052 | 0.550 ± 0.097 | 0.722 ± 0.032 | 0.779 ± 0.015 |
| 0.3 | 3.0 | 0.208 ± 0.075 | 0.494 ± 0.116 | 0.516 ± 0.059 | 0.593 ± 0.026 |

### S2.2.2 Rendering Ablation, Style Alignment, and DFT Reference

Supplementary Table 7a separates the effects of rendering basis and Pix2Pix style alignment using the 2-round product space. The vdW-sphere/Pix2Pix pathway achieved the highest micro-F1 across all tested data volumes, increasing from 0.759 ± 0.026 at n = 100 to 0.890 ± 0.007 at n = 3000. Removing Pix2Pix reduced performance to 0.690 ± 0.062 at n = 3000. DFT-based rendering generally improved with additional synthetic data but remained below vdW-sphere/Pix2Pix, reaching 0.660 ± 0.074 at n = 3000; without Pix2Pix, it reached 0.398 ± 0.035.

Supplementary Table 7b isolates the effect of Pix2Pix training-set size for the 3-round vdW-sphere pathway. Increasing the paired-image set from 50 to approximately 660 increased micro-F1 at n = 3000 from 0.702 ± 0.038 to 0.871 ± 0.016, consistent with improved transfer as the appearance-training set became more diverse.

**Supplementary Table 7. Rendering and style-alignment ablations.** (a) Effects of rendering basis and Pix2Pix style alignment in the 2-round product space. (b) Effect of Pix2Pix training-pair count for vdW-sphere rendering in the 3-round product space. Values are mean ± standard deviation; n denotes the number of synthetic YOLO training images.

| **(a) Rendering basis and style-alignment ablation (2-round product space)** | | | | | |
|---|---|---|---|---|---|
| **Rendering basis** | **Style alignment** | **n = 100** | **n = 300** | **n = 1000** | **n = 3000** |
| vdW-sphere | Pix2Pix | 0.759 ± 0.026 | 0.829 ± 0.017 | 0.857 ± 0.018 | 0.890 ± 0.007 |
| vdW-sphere | None | 0.645 ± 0.110 | 0.746 ± 0.057 | 0.712 ± 0.037 | 0.690 ± 0.062 |
| DFT | Pix2Pix | 0.186 ± 0.121 | 0.269 ± 0.103 | 0.703 ± 0.047 | 0.660 ± 0.074 |
| DFT | None | 0.069 ± 0.100 | 0.150 ± 0.132 | 0.360 ± 0.174 | 0.398 ± 0.035 |
| **(b) Pix2Pix training-pair ablation (3-round vdW-sphere pathway)** | | | | | |
| **Rendering basis** | **Pix2Pix pairs** | **n = 100** | **n = 300** | **n = 1000** | **n = 3000** |
| vdW-sphere | 50 | 0.372 ± 0.064 | 0.479 ± 0.070 | 0.634 ± 0.058 | 0.702 ± 0.038 |
| vdW-sphere | 200 | 0.430 ± 0.068 | 0.449 ± 0.120 | 0.647 ± 0.024 | 0.684 ± 0.113 |
| vdW-sphere | 400 | 0.449 ± 0.049 | 0.561 ± 0.103 | 0.667 ± 0.047 | 0.702 ± 0.035 |
| vdW-sphere | 660 | 0.737 ± 0.067 | 0.749 ± 0.062 | 0.850 ± 0.014 | 0.871 ± 0.016 |

## S2.3 Model Scaling and Computational Efficiency

### S2.3.1 Scale, Model Capacity, and Computational Cost

Effective physical sampling density, rather than field of view or pixel count alone, strongly influences transfer performance (Supplementary Fig. 8 and Supplementary Table 8). At n = 3000, the best tested combination is a 10 nm field of view represented by 256 pixels (micro-F1 = 0.869 ± 0.010), followed by the 20 nm, 512-pixel reference setting (0.839 ± 0.024); the 40 nm, 128-pixel setting yields only 0.088 ± 0.067.

**Supplementary Table 8. Joint effect of physical field of view d (nm) and pixel resolution p (pixels) on test micro-F1.**

| d (nm) | p (pixels) | n = 100 | n = 300 | n = 1000 | n = 3000 |
|---|---|---|---|---|---|
| 10 | 128 | 0.591 ± 0.024 | 0.617 ± 0.027 | 0.715 ± 0.096 | 0.762 ± 0.076 |
| 10 | 256 | 0.564 ± 0.117 | 0.713 ± 0.046 | 0.824 ± 0.043 | 0.869 ± 0.010 |
| 10 | 512 | 0.573 ± 0.069 | 0.608 ± 0.082 | 0.722 ± 0.029 | 0.832 ± 0.007 |
| 20 | 128 | 0.308 ± 0.166 | 0.474 ± 0.160 | 0.501 ± 0.098 | 0.538 ± 0.097 |
| 20 | 256 | 0.589 ± 0.097 | 0.667 ± 0.039 | 0.694 ± 0.081 | 0.732 ± 0.079 |
| 20 | 512 | 0.593 ± 0.042 | 0.672 ± 0.059 | 0.805 ± 0.032 | 0.839 ± 0.024 |
| 40 | 128 | 0.001 ± 0.002 | 0.025 ± 0.026 | 0.044 ± 0.036 | 0.088 ± 0.067 |
| 40 | 256 | 0.300 ± 0.260 | 0.448 ± 0.127 | 0.462 ± 0.106 | 0.493 ± 0.079 |
| 40 | 512 | 0.676 ± 0.032 | 0.705 ± 0.061 | 0.742 ± 0.075 | 0.731 ± 0.063 |

Increasing both synthetic-data volume and YOLO11 capacity improves test micro-F1 overall (Supplementary Table 9). With 100000 synthetic images, YOLO11x reaches the highest measured value, 0.928 ± 0.003, while YOLO11l reaches 0.921 ± 0.003. Thus, the overall data-capacity trend is positive, and the largest backbone provides the highest mean micro-F1 at each tested training-set size.

**Supplementary Table 9. Joint effect of YOLO11 backbone size and synthetic training-set size for the 3-round configuration.**

| size | n = 100 | n = 300 | n = 1000 | n = 3000 | n = 10000 | n = 30000 | n = 100000 |
|---|---|---|---|---|---|---|---|
| n | 0.520 ± 0.091 | 0.646 ± 0.070 | 0.670 ± 0.031 | 0.742 ± 0.021 | 0.864 ± 0.020 | 0.868 ± 0.022 | 0.892 ± 0.015 |
| s | 0.542 ± 0.100 | 0.734 ± 0.035 | 0.792 ± 0.010 | 0.817 ± 0.023 | 0.864 ± 0.004 | 0.879 ± 0.009 | 0.892 ± 0.007 |
| m | 0.614 ± 0.071 | 0.777 ± 0.028 | 0.834 ± 0.015 | 0.865 ± 0.017 | 0.879 ± 0.011 | 0.909 ± 0.009 | 0.912 ± 0.004 |
| l | 0.688 ± 0.080 | 0.792 ± 0.032 | 0.843 ± 0.024 | 0.869 ± 0.018 | 0.889 ± 0.011 | 0.913 ± 0.007 | 0.921 ± 0.003 |
| x | 0.720 ± 0.056 | 0.831 ± 0.026 | 0.864 ± 0.010 | 0.879 ± 0.011 | 0.908 ± 0.012 | 0.923 ± 0.005 | 0.928 ± 0.003 |

### S2.3.2 Training-Time Analysis

The complete computational timing results are reported in Supplementary Table 10, while the preprocessing and coordinate transformations used for non-reference scan formats are illustrated in Supplementary Fig. 9.

The reported computing environment used an NVIDIA GeForce RTX 4070 Ti SUPER GPU (16 GiB nominal device memory; approximately 17.2 GB in decimal units), NVIDIA driver 560.94, CUDA 12.6, cuDNN 9.10.02, Python 3.10.19, PyTorch 2.9.1+cu126, Ultralytics 8.3.247, and RDKit 2025.09.3.

**Supplementary Table 10. Computational timing for synthetic image generation, model training, and representative end-to-end workflows.**

(a) Synthetic image-generation timing.

| Statistic | Images | Time per image (s) | Generation time for 1000 images |
|---|---|---|---|
| Mean ± SD | 1000 per round | 0.065 ± 0.002 | 64.821 ± 2.397 s (1.080 ± 0.040 min) |

(b) Total model-training time for 100 epochs (min).

| YOLO11 | n = 100 | n = 300 | n = 1000 | n = 3000 | n = 10000 | n = 30000 |
|---|---|---|---|---|---|---|
| n | 2.560 | 7.690 | 25.630 | 76.900 | 256.330 | 769.000 |
| s | 2.560 | 7.680 | 25.580 | 76.750 | 255.830 | 767.500 |
| m | 3.180 | 9.530 | 31.750 | 95.250 | 317.500 | 952.500 |
| l | 4.190 | 12.570 | 41.900 | 125.700 | 419.000 | 1257.000 |
| x | 5.670 | 17.020 | 56.730 | 170.200 | 567.330 | 1702.000 |

(c) End-to-end NanoSim workflow time (min).

| Configuration | Product prediction | Image generation | Training (100 epochs) | Total workflow time |
|---|---|---|---|---|
| YOLO11l, n = 1000 | 1.000 | 1.080 | 41.900 | 43.980 |
| YOLO11l, n = 3000 | 1.000 | 3.241 | 125.700 | 129.941 |
| YOLO11l, n = 10000 | 1.000 | 10.804 | 419.000 | 430.804 |
| YOLO11x, n = 300 | 1.000 | 0.324 | 17.020 | 18.344 |
| YOLO11x, n = 1000 | 1.000 | 1.080 | 56.730 | 58.810 |
| YOLO11x, n = 3000 | 1.000 | 3.241 | 170.200 | 174.441 |
| YOLO11x, n = 30000 | 1.000 | 32.411 | 1702.000 | 1735.411 |

Image-generation values are measured over three independent 1000-image rounds. Training-time entries are reported for the fixed 100-epoch schedule in the computing environment specified above; where applicable, totals are calculated from the corresponding per-epoch timing and the 100-epoch schedule. End-to-end time is calculated as product prediction (1.000 min) plus image generation at the measured mean rate (0.064821 s per image) plus model training.

## S2.4 Experimental STM Applications

Supplementary Figures 14–18 provide additional class-resolved results beyond the representative examples presented in the main text, including molecular assignments, orientation distributions, intermolecular-distance distributions, annealing-dependent populations, and 5-round detection statistics. These observables are derived from OBB centers, orientations, class identities, and confidence scores.

### S2.4.1 Molecular Assignments and Spatial Observables

Representative STM crops paired with the corresponding molecular structures provide direct visual support for the class assignments used in the quantitative analysis, particularly for structurally related products with similar experimental contrasts (Supplementary Fig. 14).

Class-resolved orientation distributions derived from the predicted OBB angles convert individual detections into a physically interpretable system-level observable (Supplementary Fig. 15). Reporting each class separately prevents the alignment of rare species or broad distributions of abundant species from being obscured by aggregation.

Class-pair intermolecular-distance distributions derived from predicted OBB centers characterize spatial organization beyond relative abundance (Supplementary Fig. 16). Rare class pairs contain fewer observations and should therefore be interpreted cautiously.

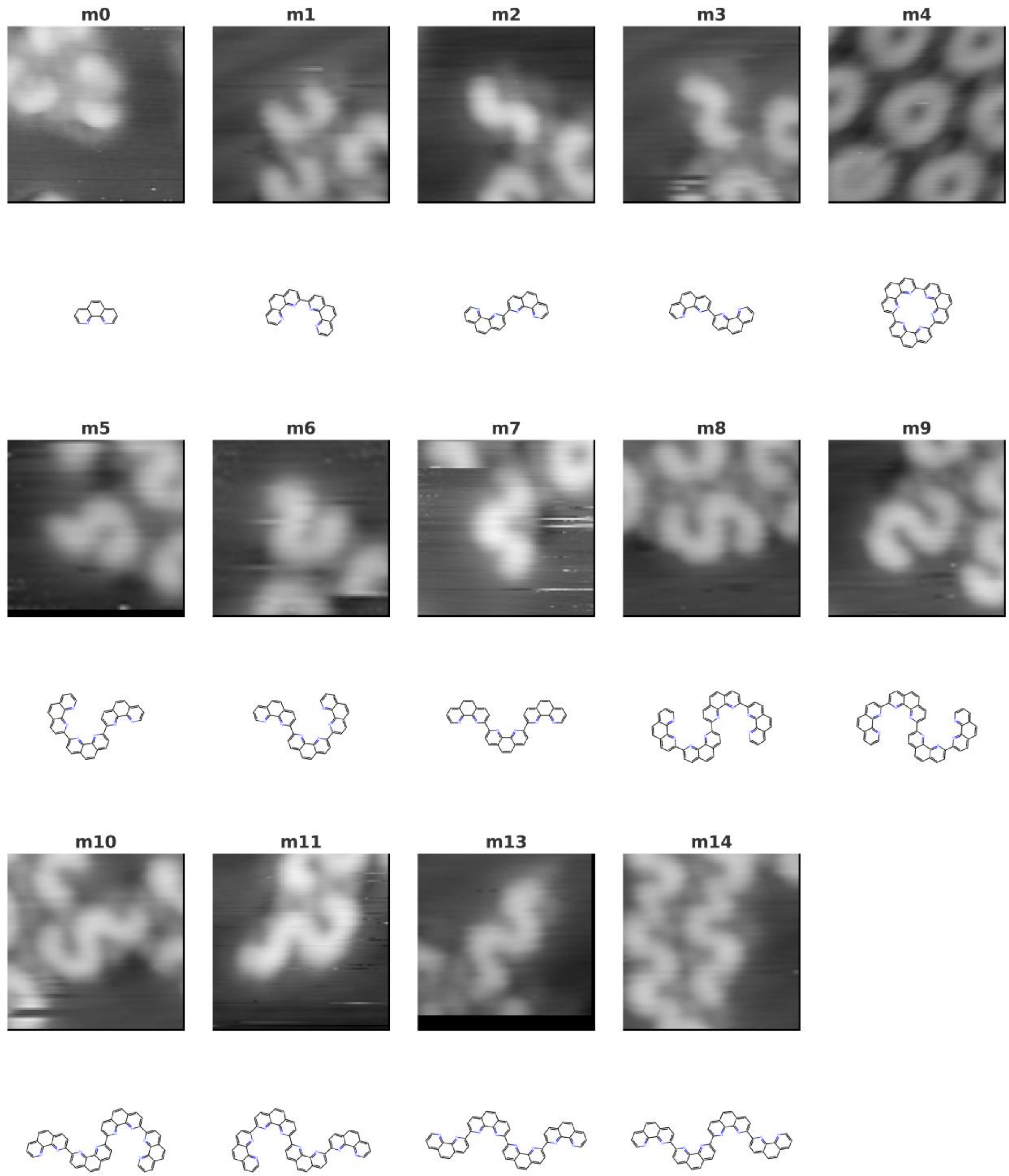

**Supplementary Figure 14. Representative STM crops and molecular assignments.** Experimental observations are paired with structures for the 14 experimentally observed classes in the 3-round BP product space (m0-m14, excluding m12).

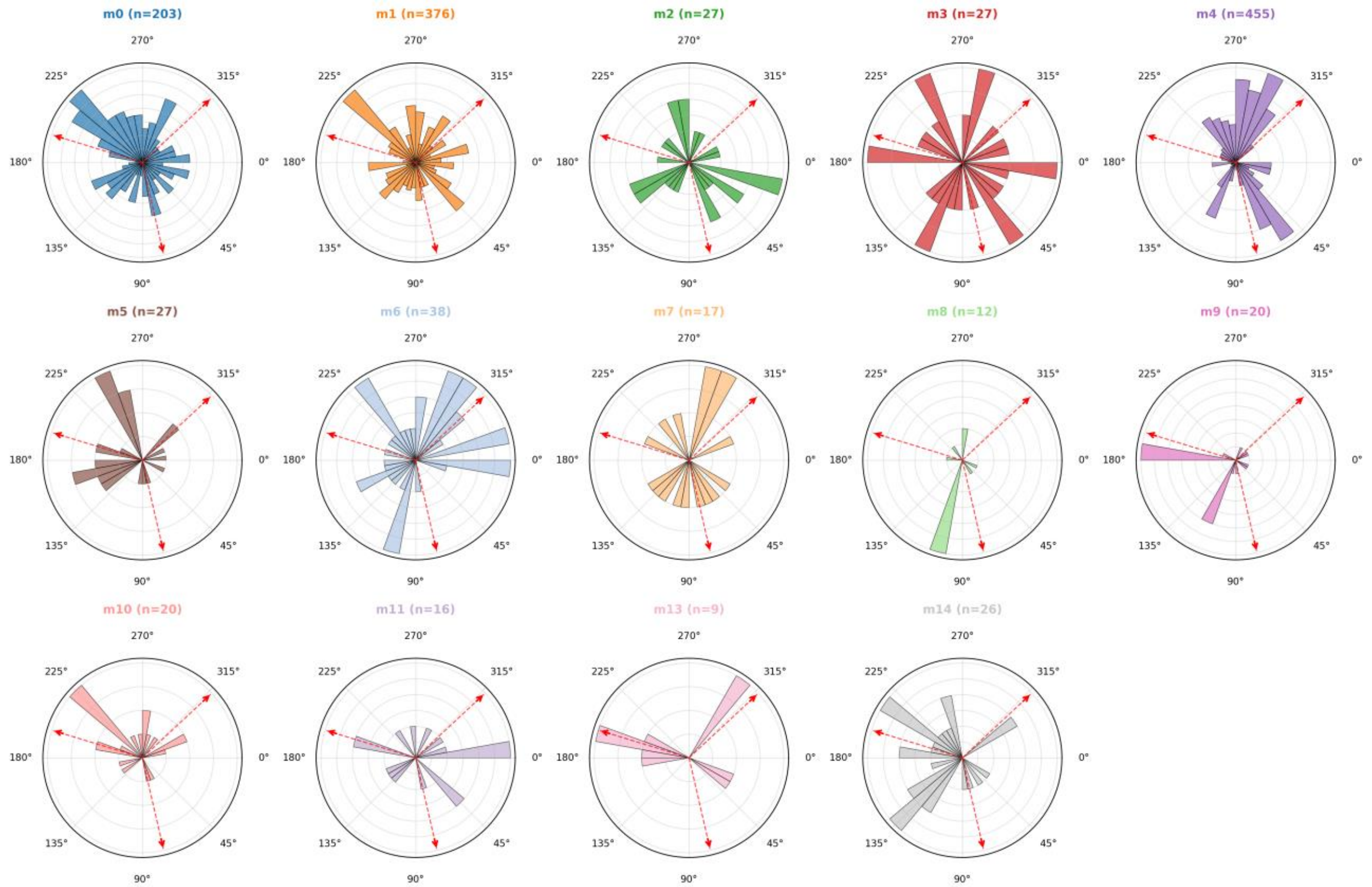

**Supplementary Figure 15. Class-resolved molecular orientation distributions.** Dashed lines indicate reference symmetry directions; panels report detection counts and circular statistics.

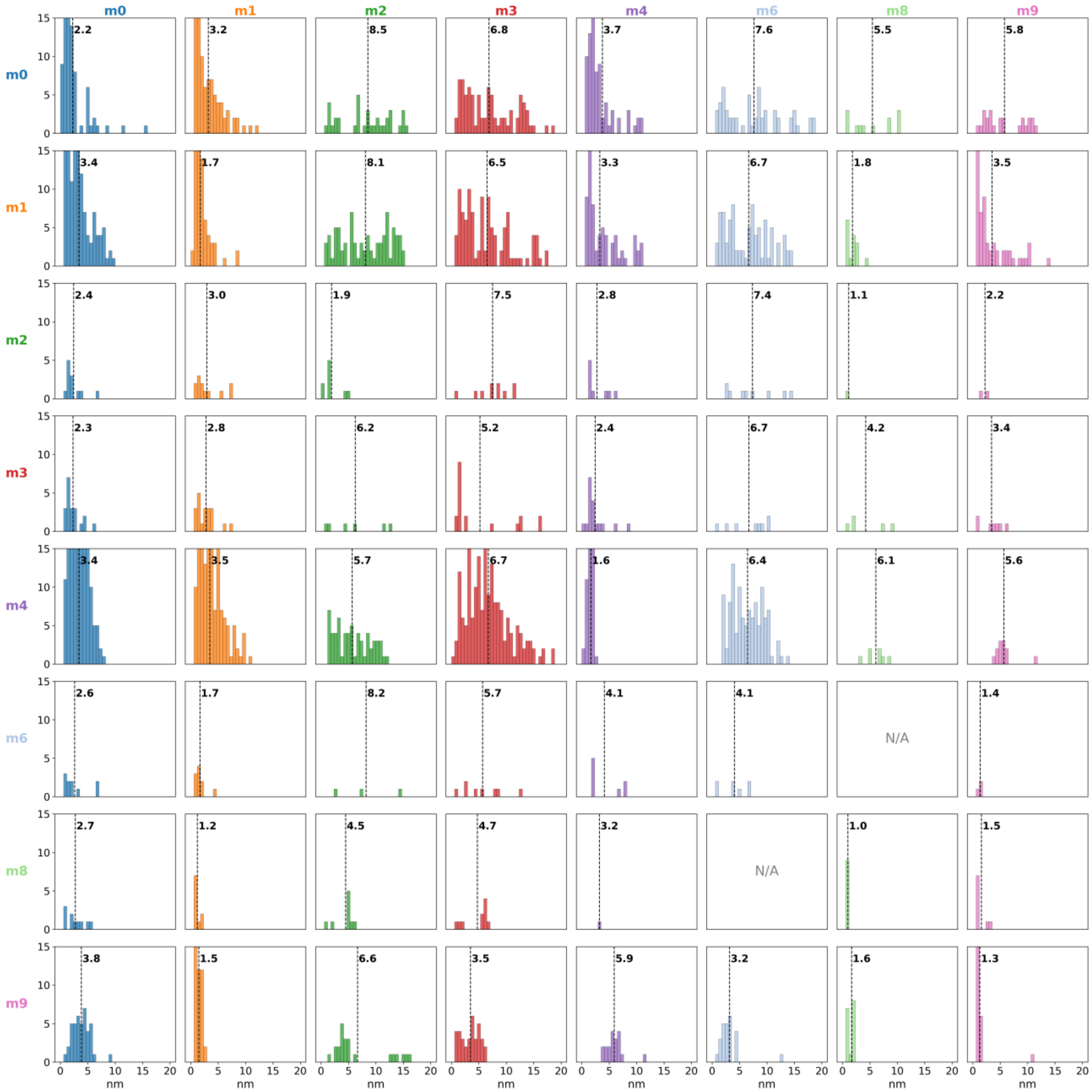

**Supplementary Figure 16. Intermolecular-distance distributions.** Diagonal and off-diagonal panels show within-class and between-class center-to-center distances, respectively.

### S2.4.2 Annealing Population-Flow Analysis

Annealing produces a pronounced redistribution from precursor-associated detections toward later products (Supplementary Fig. 17). Images were sampled randomly within each experimental stage: 20 pre-annealing images and 40 images after each of the 20- and 30-min annealing treatments were analyzed. The smaller pre-annealing sample is due to the limited number of available images at this stage. The pre-annealing set contains 613 detections across 20 images (542 m0 and 71 m1), while the 20-min and 30-min sets contain approximately 1,140 and 1,148 detections. Because the stages contain different numbers of images, stage-normalized fractions provide the more direct comparison: the m0 fraction decreases from 88% before annealing to 12% and 10% after 20 and 30 min, whereas the m4 fraction increases from 0% to 12% and 44%, respectively. These detection-based population estimates may be affected by class-dependent recognition performance and should therefore be interpreted primarily in terms of relative trends rather than molecule-resolved reaction trajectories.

The complete stage-specific image pools comprise 20 pre-annealing images, 40 images after 140 °C for 20 min, and 40 images after 140 °C for 30 min (100 images in total).

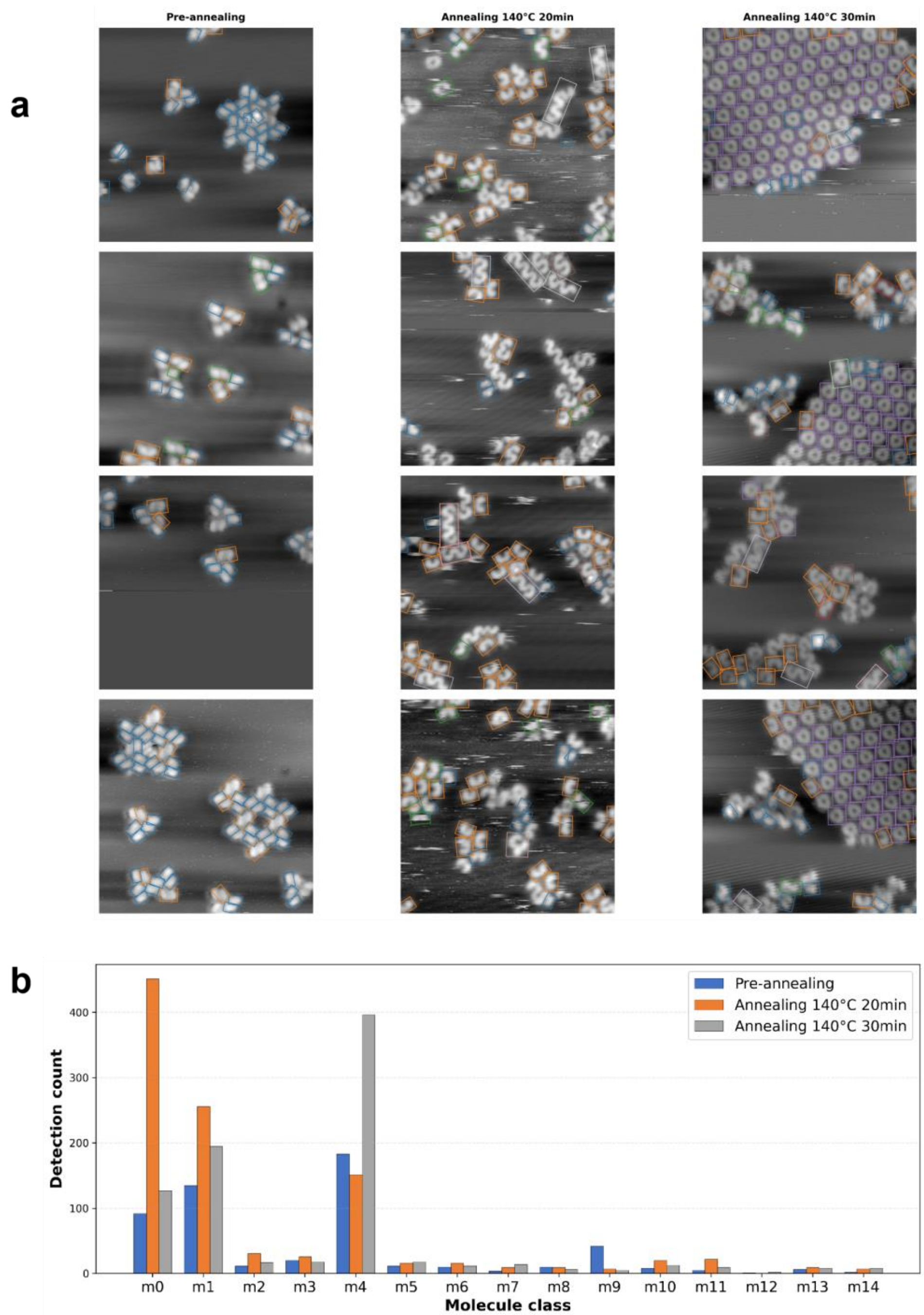

**Supplementary Figure 17. Annealing-dependent molecular populations.** (a) Representative OBB detections before annealing and after 140 °C for 20 and 30 min. (b) Class-resolved detection counts. Image scale: 20×20 nm2.

### S2.4.3 5-Round Class-Resolved Detection Statistics

Supplementary Figure 18 summarizes predictions across the enlarged 5-round label space, which contains the precursor m0 and 38 reaction products (m1-m38). Detection counts and summed confidence are shown separately for m0-m14 and m15-m38 to preserve the readability of molecular structures and rare-class outputs.

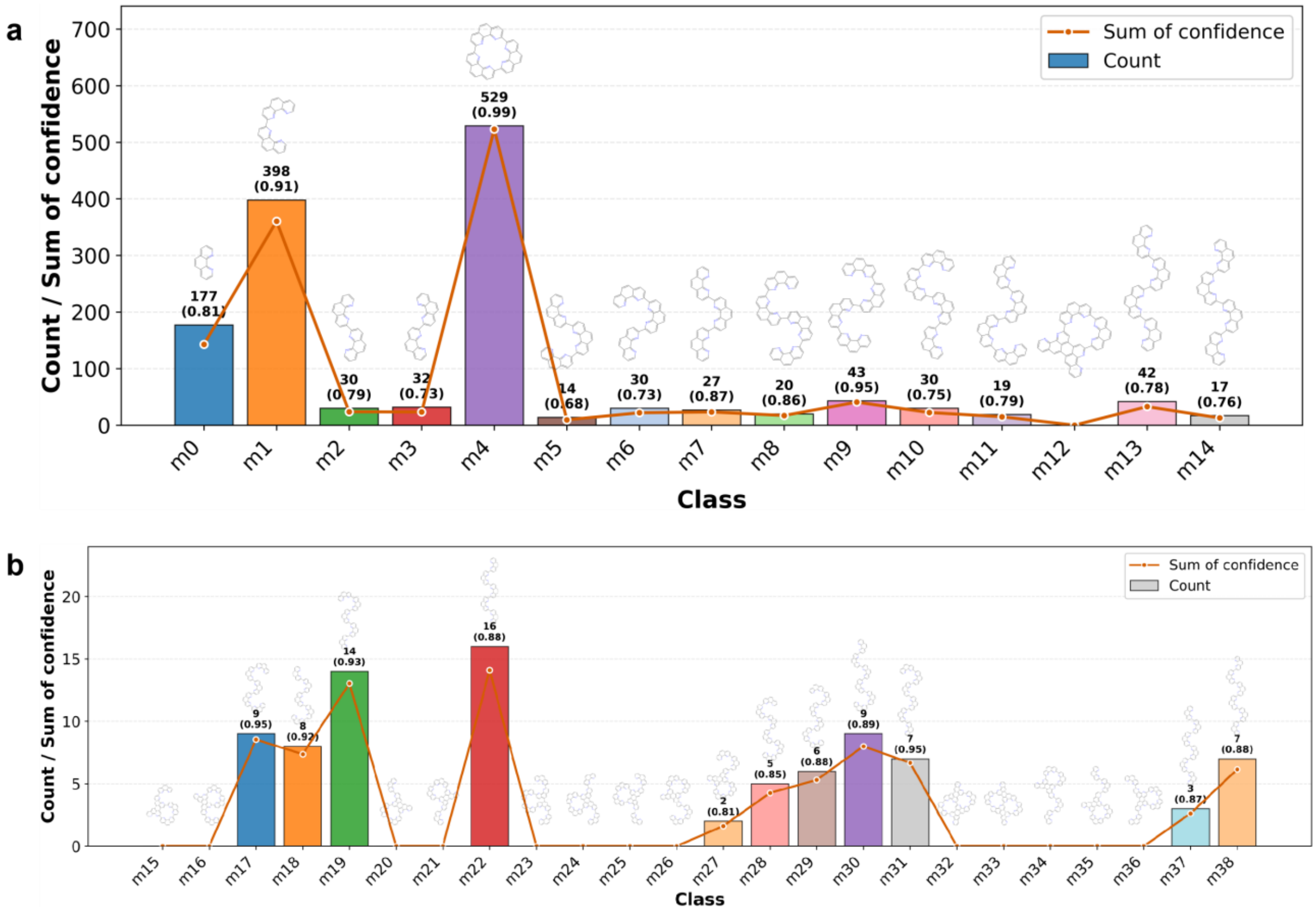

**Supplementary Figure 18. Class-resolved 5-round detection statistics.** (a) Classes m0-m14. (b) Classes m15-m38. Bars show detection counts; orange markers show summed confidence.

# S3 Model Limitations

## S3.1 Detection Failure Modes and Current Model Limitations

Supplementary Figure 19 compares predictions obtained using the 3-round and 5-round reaction-guided product spaces on the same experimental STM regions. Overall, both models detect most spatially isolated molecular features and remain effective in densely packed regions. The principal errors arise from ambiguous projected morphology, objects absent from the target product space, and closely spaced molecules whose image features overlap and become difficult to distinguish.

A representative classification error occurs when two neighboring monomers adopt an arrangement resembling the projected morphology of a dimer. In such cases, the detector may merge the two monomeric features and assign them to a dimer class. This error is associated with the similarity between the combined STM appearance of the adjacent monomers and the learned appearance of a coupled product. Objects outside the reaction-guided product space may remain undetected, be assigned to a visually similar known class, or be represented by multiple partial detections.

Expanding the reaction-guided product space from 3 to 5 rounds improves the recognition of several larger and elongated oligomer-like features. The 5-round product space contains a broader range of higher-order products and therefore provides candidate classes that more closely match some experimentally observed oligomers. Consequently, several features that are missed, fragmented, or incorrectly assigned by the 3-round model are detected more consistently by the 5-round model. Structures that remain outside the expanded product space can still be missed or assigned to the closest available class.

Both models retain good localization performance in densely packed molecular regions. Orientation errors are more common for molecules with highly symmetric projected shapes, because multiple in-plane orientations can produce nearly indistinguishable STM appearances. Consequently, a molecule may be correctly localized and classified even when the predicted oriented-bounding-box angle differs from the visually assigned orientation.

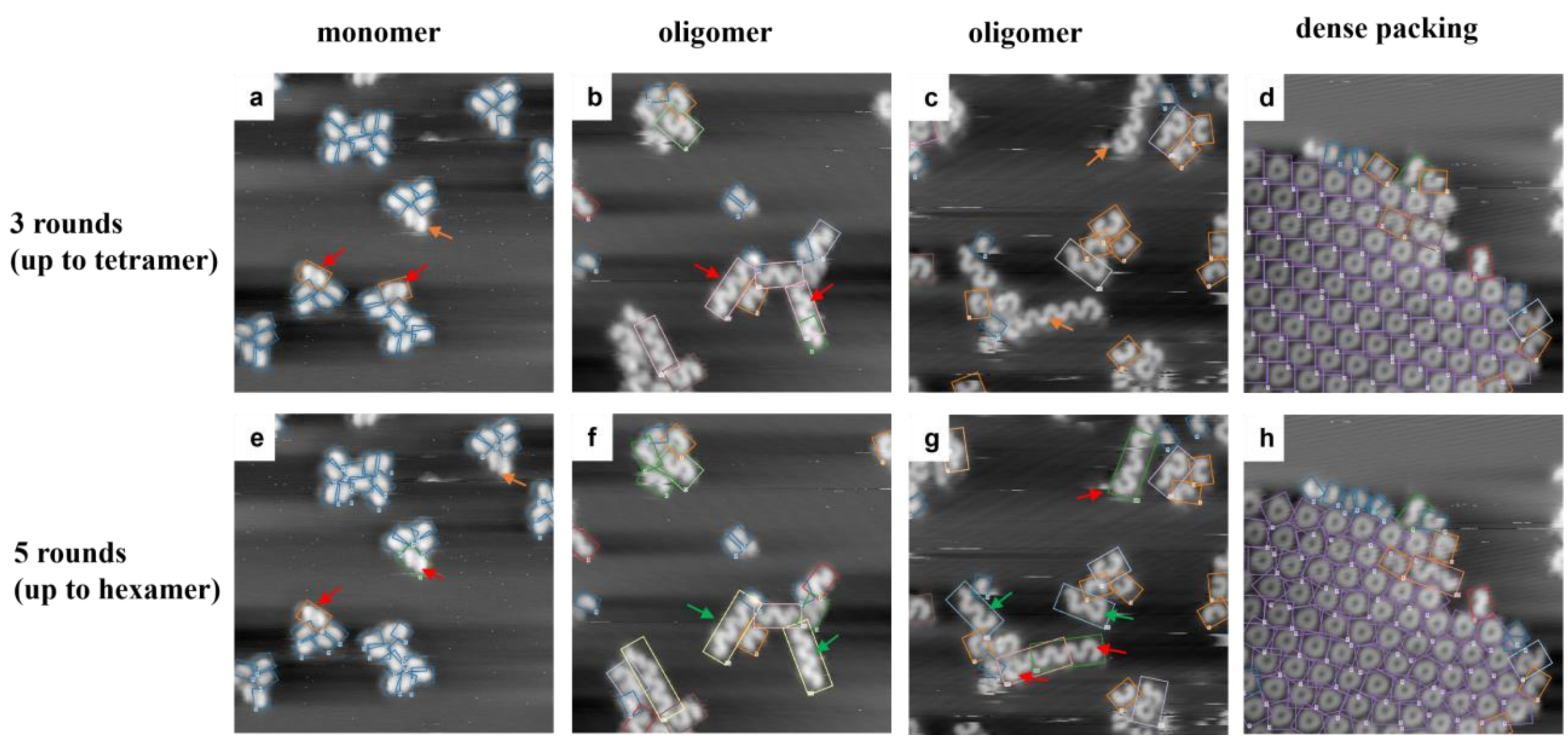

**Supplementary Figure 19. Detection results from 3-round and 5-round product-space models.** Panels (a-d) and (e-h) show corresponding STM regions for the 3-round and 5-round models, respectively. Orange arrows mark missed detections, red arrows mark erroneous or ambiguous assignments, and green arrows mark detections improved with the 5-round model. Image scale: 20×20 nm2.

## S3.2 Limitations of the Simulation Framework

NanoSim's detection performance depends on the correspondence between the simulated and experimental product, scene, and image spaces. At the Physical layer, reaction-guided construction restricts the candidate classes to products predicted within the selected reaction sites and reaction depth. This restriction keeps the recognition problem manageable but may exclude products that are present experimentally. Such structures may consequently be missed or assigned to the closest known class by the current closed-set detector. Expanding the product space from 3 to 5 rounds improves the recognition of some larger molecular features (Supplementary Fig. 19), but also increases class complexity and training-data requirements.

At the Behavior layer, synthetic scenes are generated using predefined distributions of class abundance, molecular density, intermolecular separation, and orientation. Differences between these assumptions and the experimental surface organization may cause distribution shifts and reduce transfer performance. Supplementary Figure 13 and Supplementary Table 4 show that the molecular-sampling weight affects class balance and detection performance, whereas Supplementary Tables 5–6demonstrate sensitivity to orientation, density, and intermolecular separation. Scene sampling should therefore cover plausible experimental variation rather than rely on a single configuration.

At the Rendering layer, Pix2Pix converts geometry-based representations into STM-like images. This transformation learns aggregate appearance signatures associated with the tip state, substrate, imaging conditions, instrumental response, and noise, but the finite paired training set may not represent all experimental appearances. The dependence on the number of paired images in Supplementary Table 7 suggests that insufficient rendering-domain coverage can limit transfer. Because mismatches introduced at the three layers may accumulate, the resulting molecular assignments and quantitative statistics should be interpreted within the distributions represented during training, rather than as an exhaustive description of all surface products or reaction pathways.

# S4 Transfer to Additional Molecular Systems

To examine transferability beyond the BP system, we applied the simulation-assisted framework to two chemically distinct molecular systems, DBAP and porphine (Supplementary Fig. 20).

## S4.1 DBAP Molecular System

Supplementary Figure 20a–cdemonstrates transfer of the workflow to the DBAP system.[13] The reaction-guided product space contains the DBAP precursor (mol_0) and three coupled candidates, among which mol_1 corresponds to the experimentally reported diaza-HBC product formed through surface-assisted homocoupling and cyclodehydrogenation (Supplementary Fig. 20a). Before annealing, the model qualitatively recovers most visible precursor molecules (Supplementary Fig. 20b); after annealing, it identifies the mol_1 product together with residual precursor molecules (Supplementary Fig. 20c).

## S4.2 Porphine Molecular System

Supplementary Figure 20d–f illustrates transfer of the workflow to the porphine system. Previous studies have identified several porphine dimerization motifs, including the symmetric triply fused β-β/meso-meso/β-β motif and the doubly fused meso-β/β-meso and β-β/meso-β motifs.[14,15] The current reaction-guided product space contains only the triply fused dimer (mol_1), whereas the two alternative motifs are not yet represented (Supplementary Fig. 20d). Before annealing, the model recovers most visible monomers but also produces several apparent dimer false positives (Supplementary Fig. 20e). After annealing, three dimer-like features are detected, together with one additional apparent false positive (Supplementary Fig. 20f). These results therefore demonstrate qualitative, but not yet exhaustive, transfer to the porphine system.

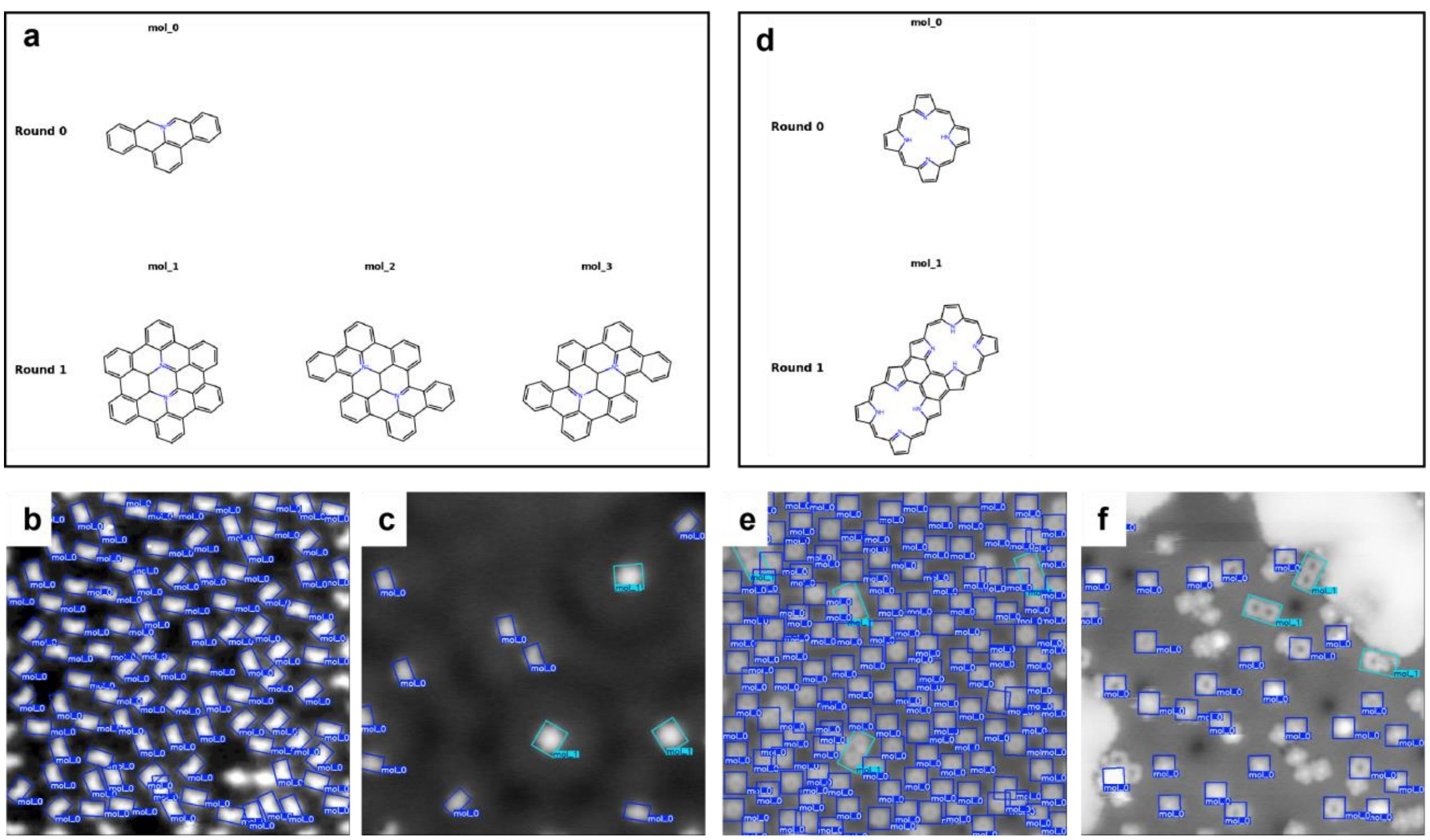

**Supplementary Figure 20. Transfer to DBAP and porphine molecular systems. (a) DBAP precursor and predicted coupled products; (b, c) representative DBAP detections. (d) Porphine precursor and predicted dimer; (e, f) representative porphine detections. Image scale: 20×20 nm2.**

# Supplementary References